\documentclass{easychair}

\usepackage{long2}
\usepackage{pdflscape}
\usepackage{longtable}
\usepackage{cite}
\usepackage{epstopdf}
\usepackage{dingbat}
\usepackage{pifont}
\newcommand{\cmark}{\ding{51}}%
\newcommand{\xmark}{\ding{55}}%

\begin{document}

%
\title{A Survey on the Security of Pervasive Online Social Networks (POSNs)}

%
\titlerunning{POSNs: Security}

\volumeinfo
	{\emph{Journal of Internet Services and Information Security (JISIS)}}
	{8}                         
	{2  (May 2018)}                         
	{48}                         

%
\author{\\
Takshi Gupta, Gaurav Choudhary, and Vishal Sharma\thanks{Corresponding author: Dept of Information Security Engineering, Soonchunhyang University, 22 Soonchunhyang-ro, Shinchang-myeon, Asan-si, Choongchungnam-do, Republic of Korea 31538, Tel: +82-(0)41-530-3099}\\
Dept of Information Security Engineering, Soonchunhyang University\\
takshi\_gupta2012@hotmail.com, gauravchoudhary7777@gmail.com,\\
vishal\_sharma2012@hotmail.com
}

%
\authorrunning{Gupta \emph{et al.}}

\maketitle

%
\begin{abstract}
\noindent
Pervasive Online Social Networks (POSNs) are the extensions of Online Social Networks (OSNs) which facilitate connectivity irrespective of the domain and properties of users. POSNs have been accumulated with the convergence of a plethora of social networking platforms with a motivation of bridging their gap. Over the last decade, OSNs have visually perceived an altogether tremendous amount of advancement in terms of the number of users as well as technology enablers. A single OSN is the property of an organization, which ascertains smooth functioning of its accommodations for providing a quality experience to their users. However, with POSNs, multiple OSNs have coalesced through communities, circles, or only properties, which make service-provisioning tedious and arduous to sustain. Especially, challenges become rigorous when the focus is on the security perspective of cross-platform OSNs, which are an integral part of POSNs. Thus, it is of utmost paramountcy to highlight such a requirement and understand the current situation while discussing the available state-of-the-art. With the modernization of OSNs and convergence towards POSNs, it is compulsory to understand the impact and reach of current solutions for enhancing the security of users as well as associated services. This survey understands this requisite and fixates on different sets of studies presented over the last few years and surveys them for their applicability to POSNs. There is a limited amount of content available for the security of POSNs. However, being an extension to general OSNs, solutions applicable to OSNs are withal included to understand the practicality for POSNs. Moreover, this survey additionally includes content cognate to trust management and anomaly detection in POSNs.  In integration, a broad classification is additionally presented for each category with a tabular comparison. At last, certain future challenges, open issues, and research goals are presented, which can be focused by leading or upcoming researchers while emphasizing the security of POSNs.
\end{abstract}
\section{Introduction}
\begin{figure}[ht!]
\centering
\includegraphics[width=340px]{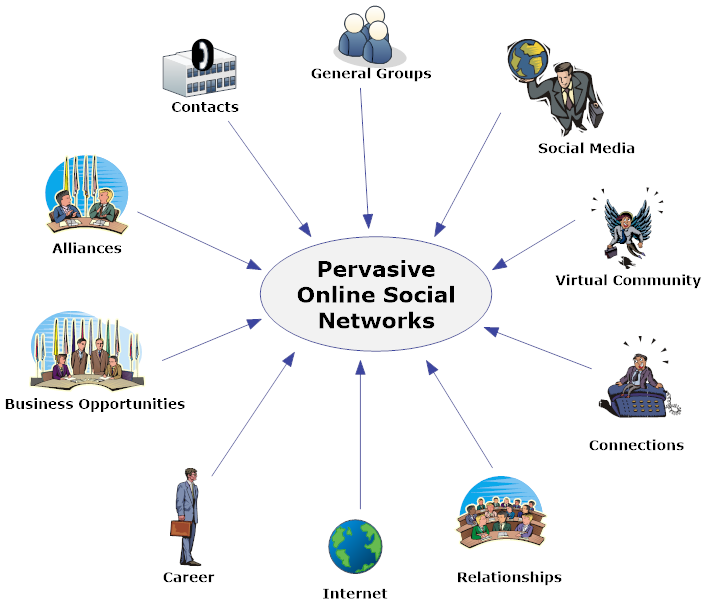}
\caption{Applications of pervasive online social networks.}
\label{fig2}
\end{figure}

Connecting users irrespective of their domain and online platform is studied under Pervasive Online Social Networks (POSNs). POSNs help to facilitate pervasive social networking by eliminating the resource and architectural boundaries of the OSN platforms. POSNs allow inter-connectivity amongst vast domains of OSN applications like contact-building, news-sharing, business modeling, online content-making, career build-ups, etc, as shown in Figure~\ref{fig2}. In general, POSNs comprise of a large set of communities which belong to different social network platforms and are combined together on the basis of common interests or roles~\cite{sharma2017computational}~\cite{shehnepoor2017netspam}. Apart from these, POSNs also accommodate vehicular social networks which involve dynamic components as a social networking entity, such as smart cars, robots, drones, or even autonomous vehicles~\cite{shin2017secure}~\cite{zhang2017social}~\cite{sharma2017saca}.

In spite of the tremendous amount of facilities and advantages, POSNs suffer from critical issues related to security, privacy, trust and anomaly detections~\cite{dincelli2017can}~\cite{gavrilova2017emerging}. POSNs require sufficiently efficient, but low complex solutions to identify and mitigate the threats associated with the smooth functioning over platforms that support its applications. Trust helps to formalize relationships amongst the entities involved in POSNs while leveraging on the popular solutions like reputation building, Peer to Peer, or Peer to Multi-Peer connectivity~\cite{sun2017user}~\cite{perera2017wireless}. However, the success of such requirements depends on the level of security accommodation supported by policies for POSNs.

In addition, operational profilers can be used to determine the roles users as well as for the classification of communities. Successful classification of communities supports easier management of POSNs as well as helps to regulate the common operations without any interruptions. Detection of faulty users in POSNs can be carried either through the involved platforms or multiple platforms can depend on the third-party evaluators. However, these solutions are dependent on the privacy policies as well as the personalized settings of each involved community or a user.

Vulnerabilities in POSNs tend to gain access over the unsecured wireless links and identify systems with failed security patches and software-prone to different attacks~\cite{zhang2010privacy}~\cite{al2009threats}. Further, attacks on POSNs expose the passwords, device-information as well as feed users with malicious and undesirable contents. Threats in POSNs can be due to malicious programs such as viruses, trojans, malware, worms, and botnets. Especially, over the last decade, adware has been the leading cause of attacks on most of the OSN platforms and has potential to exploit users in POSNs. Alongside, Denial of Service (DoS), Distributed Denial of Service Attacks (DDoS), Data theft, Sybil attacks, Blackhole attack, Greyhole attack, and zero-hour attacks are other known attacks that can target POSNs~\cite{wood2002denial,sharma2017consensus,irani2011reverse,gao2011security,sharma2018framework}.

\begin{figure}[ht!]
\centering
\includegraphics[width=340px]{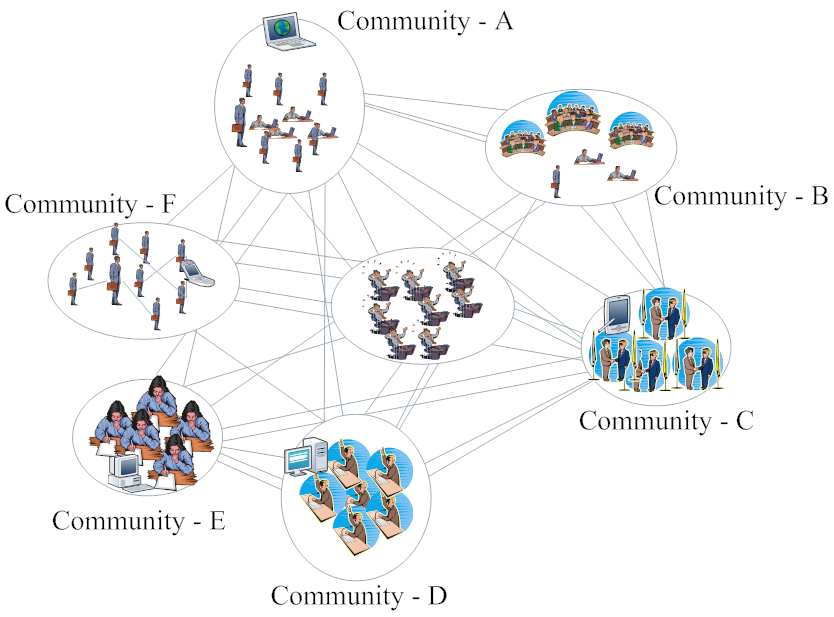}
\caption{An exemplary illustration of pervasive online social networks through multiple communities.}
\label{fig1}
\end{figure}

Intentional user-access privileges and the unwanted inclusion of service-requests pose a significant security threat in POSNs. Management of users, marking of communities, identification of platforms, accessibilities to cross-platform services and data lookups should be prioritized and limited accessibility should be given to a regular user in POSN, as shown in Figure~\ref{fig1}~\cite{sharma2017managing}. Such a management can help to reduce the severity of an attack, which might be launched in future. Frequent evaluations should be conducted to check the privileges of maintenance as well as root users. Machine learning and Artificial Intelligence (AI) can be used to develop software that can periodically check user accounts and can help to clean up inactive or largely requesting accounts on cross-platform POSNs.

Moreover, Internet security solutions, strong authentication, and novel key agreement protocols can also help to formalize solutions for preventing any known cyber attack on POSNs~\cite{guan2017extension}~\cite{weng2016peer}. Such solutions can be supported by techniques like packet filtering, stateful packet inspection, firewalls, security tokens, or even through authentication protocols like Extensible Authentication Protocol (EAP), Identity Authentication Protocol (IAP), Password Authentication Protocol (PAP), Host Identity Protocol (HIP), Secure Remote Password protocol (SRP), Challenge-Handshake Authentication Protocol (CHAP), etc~\cite{nakhjiri2005aaa}~\cite{forouzan2006data}.
\label{sect:introduction}

\begin{figure}[ht!]
\centering
\includegraphics[width=340px]{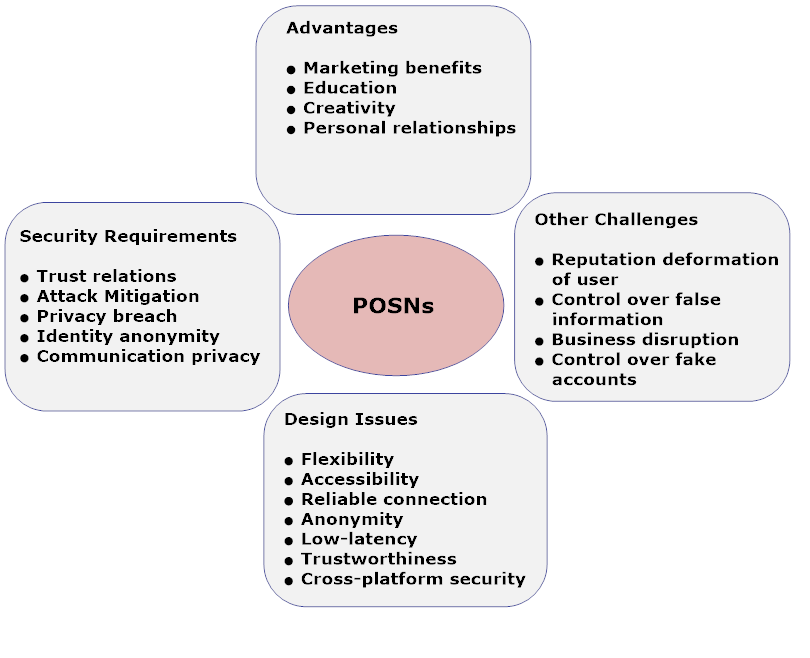}
\caption{Advantages and challenges associated with the implementation of POSNs.}
\label{fig3}
\end{figure}

\subsection{Advantages of POSNs}
Online social communities emerge exponentially over the POSNs. Social networking platforms like Facebook, Twitter, etc, empower the users and increase their interests to exchange information and resources. Some of the advantages of POSNs, as shown in Figure~\ref{fig3}, are discussed below.
\begin{itemize}
\item \textbf{Marketing benefits :} POSNs plays an important role in business advertisements. Nowadays, companies, artists, and musicians can reach an impossibly large and diverse amount of people using cross-platform services. Social media can be used as of promotional tool for marketing of their products. Targeting specialized community and showcasing results of interest are major advantages of POSNs.
\item \textbf{Education :} POSNs can be regarded as a platform allowing academicians and industry-researchers to communicate and collaborate on their requirements.
\item \textbf{Creativity :} These platforms boost the users to perform creative activities with their friends and communities. These platforms are open to think and share their ideas with different users.
\item \textbf{Personal relationships :} Social relationships and personal relationships can be maintained through POSNs, and information/profiles on one can be shared with other platforms without recreational overheads.
\end{itemize}

\subsection{Challenges of POSNs}
POSNs have considerable constraints in terms of security requirements as discussed below:
\begin{itemize}
\item \textbf{Developing secure and reliable relationship model :} POSNs rely on relationship models for interaction between the users and the communities. The social relationships involve different paradigms in terms of strength and scope of connections. The relationship models consist of the type of relationship, the strength of trust and the intensity of interactions. An effective relationship model can help to enhance the reliability and scalability of POSNs. Moreover, a relationship model helps to maintain a tradeoff between the complexity and accuracy of search and connection establishment amongst different communities~\cite{waters2000secure}.
\item \textbf{Reputation defamation of user :} Reputation is a very important asset in term of values and credibility. The reputation is directly proportional to the trust of POSNs. If the information is compromised then it destroys the credibility and reputation of a person, institution, organization, social group, or nation which exists in POSNs~\cite{carter2002reputation}. Therefore mitigation of this risk is another primary concern for POSNs.
\item \textbf{Control over false information :} The information registered over POSNs should be legitimate and it should be verified from the perspective of security. The false information can be used to defame the person or any organization. The spreading false information is very critical over POSNs because it leads to conflicts or any critical loss of any origination or any person. Therefore to resolve this issue, POSNs require protocols and policies to be added for controlling above discussed issues.
\item \textbf{Business disruption :} The organizational communities are important pillars of POSNs. The information to advertise the business of any organization leads to the misuses of assets if the false or forged information is used for advertisement. Therefore, privacy against the shared information is a considerable challenge to resolve for POSNs.
\item \textbf{Control over fake accounts :} Recently, forging and duplication of accounts have rapidly increased over social media. The duplicate accounts, which are used by the attacker, spread lots of false information through their fake accounts and should be avoided in POSNs.
\end{itemize}

\subsection{Design Issues of POSNs}
The traditional goals of POSNs such as usability and sociability are influenced by the functionalities and designing of cross-platform services~\cite{preece2005online}~\cite{gross2005information}. But there exist certain design issues which have to be focused while implementing POSNs. These include,
\begin{itemize}
\item \textbf{Flexibility :} POSNs utilize centralized web server and all the information is to be processed within the design consideration of the underlaid platform. POSNs provide all the functionalities, like storage, maintenance, and access to OSN services. Therefore, the flexibility of POSNs is required to provide large computation and handling of multi-users at a single instance and at low-complexity.

\item \textbf{Accessibility :} Many communities are connected through POSNs for accessing diversified services. The accessibility helps to maintain a check on the information and the content which are to be shared among the users of POSNs.

\item \textbf{Reliable connection :} POSNs are defined over connected graphs where the user can access and share their relationships with other. The connected graphs i.e. reliable connection, are required to maintain communication between different communities, which should be facilitated through reliable connectivity and data exchanges.

\item \textbf{Anonymity :} The protection of a user’s identity helps anonymity. The visibility of a user’s profile reveals lots of personal information. Therefore, masking of the identity or pseudo-identity is required to hide the user’s personal information. This also prevents stalking and identity thefts.

\item \textbf{Low-latency :} With an increased use of POSNs, low latency solutions are required for authentication and information access by third-party applications through APIs.

\item \textbf{Trustworthiness :} In POSNs, two different users communicate with each other, therefore, the evaluation of their trustworthiness along a certain transfer path is mandatory. The trust map can be used for such a requirement that represents the connectivity of users as a result of their trust values and overall trustworthiness.

\item \textbf{Cross-platform security :} Cross-platform OSNs may allow the same person to have multiple accounts on multiple platforms. Therefore, the cross-platform availability and reputation mechanism are mandatory in POSNs. Web services play a significant role in the interoperability of cross platforms. Therefore, cross-platform services with high security are major issues in POSNs.
\end{itemize}

\subsection{Security requirements in POSNs}
POSNs store a huge amount of critical information about users and their private conversations. The cyber-attacks like identity thefts, stalking, cyber bullies, etc, are responsible for information leakage~\cite{ellison2007benefits}~\cite{cutillo2009safebook}. Therefore, the effective and flexible security mechanisms are required for the safety of POSNs as expressed below:
\begin{itemize}
\item \textbf{Trust relations :} Trusted security mechanisms help to protect social graphs in POSNs. These security mechanisms are needed to protect POSNs and mitigate different kinds of attacks on the formed online social graphs.

\item \textbf{Attack mitigation :} A cyber attack is a harmful activity performed by the attacker to gain information relevant to the users. The information and privacy are compromised because of such type of attacks. Therefore, attack mitigation schemes over POSNs are required. The attacks can be reduced by applying some authentication and authorization policies along with cryptographic mechanisms.

\item \textbf{Privacy :}  POSNs consist of large information like user’s account details, communication messages, information regarding service provider, third-party applications, and advertisers. Users' sharing diversity and specificity of the personal information over POSNs will increase the risk for cyber and physical attacks. Therefore, fine-grained privacy setting is required in POSNs.

\item \textbf{Identity anonymity :}  The identity of users must be hidden from each other during messages exchange. Some pseudonymous-based identity hidden mechanisms are required to hide the original identity of the user. The random identifiers are the best solution for hiding public identity of users.
\end{itemize}
Despite the above-discussed security requirements, there are certain other security policies that are required to be incorporated in POSNs.

\subsection{Technology enablers for POSNs}
\label{sect: Reach and Technological Advancements}
POSNs aim at connecting users from different platforms of social media through a common interface, which enables the formation of communities. Despite being advantageous for general online social networks, socializing everyone is tedious and requires considerable efforts in devising solutions which can form a common platform for users from different online social networking platforms. Such a requirement makes it difficult for the security providers and demand extensive solutions which can be easily applied to POSNs and at a low cost of operations.

Existing software and technological solutions can help to overcome the security requirements of POSNs without interrupting its services. However, the subsisting technologies should be improvised to particularly target the applications associated with POSNs. An efficient solution not only helps to determine the responsiveness of a user in POSNs but also supports the classification of communities, detection of anomalous users as well as scanning of vulnerabilities and maintenance of security policies, like confidentiality, integrity, authentication, availability. Moreover, these also facilities the mitigation of cyber threats and prevent the POSNs platform from cyber bullies. To further extend the understandings on the existing solutions and technological advancements, a comparison between various technology enablers for POSNs is presented in Table~\ref{Table4}.
\subsection{Structure of the survey}
\label{sect:Scope and structure of the survey}
At first, the survey covered the basics of POSNs and various technology enablers. Section 2 presents a comparison between the proposed and the existing surveys. Section 3 presents the study on the security and its classification for POSNs. Trust management is covered in Section 4 followed by classification of anomaly detection in Section 5. Discussions and open issues are presented in Section 6. Literature classification and categorization are presented in Section 7. Finally, Section 8 concludes the paper.



\begin{table}[!ht]
\fontsize{9}{10}\selectfont
\centering
\caption{Key technology enablers for security in POSNs.}
\label{Table4}
\begin{tabular}{|c|c|c|}
\hline
\textbf{Security Technology}                                                                                         & \textbf{Property}                                                                                                                                             & \textbf{Advantages}                                                                                                                                           \\ \hline
Firewall and anti-virus \cite{fire2014online}                                                                                              & Detect threats                                                                                                                                                & \begin{tabular}[c]{@{}c@{}}Protecting computers and \\ devices against threats such\\ as malware, clickjacking,\\ and phishing attacks\end{tabular}           \\ \hline
\begin{tabular}[c]{@{}c@{}}Intrusion prevention \\ security technology \cite{fire2014online}\end{tabular}                                  & Detect threats                                                                                                                                                & \begin{tabular}[c]{@{}c@{}}Protections against broader and \\ more sophisticated attack spectrum\end{tabular}                                                \\ \hline
Web App Firewalls(WAF) \cite{razzaq2013cyber}                                                                                               & \begin{tabular}[c]{@{}c@{}}Protect web applications \\ from web exploits\end{tabular}                                                                         & \begin{tabular}[c]{@{}c@{}}Control traffic to allow or \\ block web applications\end{tabular}                                                                 \\ \hline
Deception \cite{howard2008analyzing}                                                                                                            & \begin{tabular}[c]{@{}c@{}}Create fake vulnerabilities, \\ systems, shares cookies and\\ find the attacker\end{tabular}                                       & \begin{tabular}[c]{@{}c@{}}Suitable for network,  application, \\ endpoint, and data\end{tabular}                                                             \\ \hline
Machine learning security \cite{wang2010detecting}                                                                                           & Find anomalous behavior                                                                                                                                      & \begin{tabular}[c]{@{}c@{}}Provides protections against advanced \\ persistent threats\end{tabular}                                                          \\ \hline
\begin{tabular}[c]{@{}c@{}}Cloud workload protection \\ platforms \cite{pallis2011online} \end{tabular}                                       & Single management console                                                                                                                                     & Applies security policy                                                                                                                                       \\ \hline
\begin{tabular}[c]{@{}c@{}}Network Traffic Analysis \cite{shalita2018routing} \\ (NTA)\end{tabular}                                            & \begin{tabular}[c]{@{}c@{}}Monitor network traffic,\\ flows, connections\end{tabular}                                                                         & Finds malicious intent                                                                                                                                        \\ \hline
\begin{tabular}[c]{@{}c@{}}Cloud Access Security \\ Brokers (CASBs)  \cite{poell2017social} \end{tabular}                                     & \begin{tabular}[c]{@{}c@{}}Single point of control \\ over multiple cloud services\end{tabular}                                                               & Address gaps                                                                                                                                                  \\ \hline
\begin{tabular}[c]{@{}c@{}}Software-Defined \\ Perimeters  (SDPs) \cite{cui2018ddse} \end{tabular}                                       & \begin{tabular}[c]{@{}c@{}}Logical set of disparate,\\ network-connected participants\end{tabular}                                                            & \begin{tabular}[c]{@{}c@{}}Hide the public visibility and reducing \\ the surface area for attack\end{tabular}                                                \\ \hline
\begin{tabular}[c]{@{}c@{}}Endpoint Detection and \\ Response (EDR)\cite{karamani2018improving} \end{tabular}                                     & \begin{tabular}[c]{@{}c@{}}EDR tools record numerous \\ endpoint and network events, \\ and store information in the\\ centralized database\end{tabular}      & Quickly in response                                                                                                                                           \\ \hline
Remote browser \cite{jakobson2014collaborative}                                                                                                       & \begin{tabular}[c]{@{}c@{}}Browser session from a \\ browser server\end{tabular}                                                                              & \begin{tabular}[c]{@{}c@{}}Detect malware delivered  via email, \\ URLs or malicious  websites.\end{tabular}                                                  \\ \hline
\begin{tabular}[c]{@{}c@{}}User and Entity Behavioural \\ Analytics (UEBA)\cite{acquisti2015privacy} \end{tabular}                              & \begin{tabular}[c]{@{}c@{}}User-centric analytics \\ of user behaviour\end{tabular}                                                                           & \begin{tabular}[c]{@{}c@{}}More accurate and threat \\ detection more effective\end{tabular}                                                                  \\ \hline
DevSecOps \cite{myrbakken2017devsecops}                                                                                                           & \begin{tabular}[c]{@{}c@{}}Use scripts, recipes, blueprints,\\ and templates to the underlying\\ the configuration of security \\ infrastructure\end{tabular} & \begin{tabular}[c]{@{}c@{}}Automatic security scanning \\ for vulnerabilities\end{tabular}                                                                    \\ \hline
\begin{tabular}[c]{@{}c@{}}Intelligence-driven security \\ Operations centre orchestration\\ solutions \cite{tounsi2017survey} \end{tabular} & Events-based monitoring.                                                                                                                                      & \begin{tabular}[c]{@{}c@{}}Used to inform every \\ aspect of security operations\end{tabular}                                                                 \\ \hline
Pervasive trust services \cite{ifinedo2016applying}                                                                                          & \begin{tabular}[c]{@{}c@{}}Designed to scale and support\\ the needs of devices with\\ limited processing capability\end{tabular}                             & \begin{tabular}[c]{@{}c@{}}Trust services include secure \\ provisioning, data integrity, confidentiality, \\ device identity and authentication\end{tabular} \\ \hline
\begin{tabular}[c]{@{}c@{}}Blockchain principles to\\ be applied to data security \cite{bahri2018decentralized}\end{tabular}                       & \begin{tabular}[c]{@{}c@{}}Blockchain has the potential \\ to be a major leap forward for \\ securing sensitive information\end{tabular}                      & \begin{tabular}[c]{@{}c@{}}Mitigate the increasing number of cyber \\ threats to data\end{tabular}                                                            \\ \hline
Data loss prevention \cite{he2018latent}                                                                                                & \begin{tabular}[c]{@{}c@{}}Provides encryption and \\ tokenization\end{tabular}                                                                               & \begin{tabular}[c]{@{}c@{}}Protect data down to field \\ and subfield level\end{tabular}                                                                      \\ \hline
\end{tabular}
\end{table}

\begin{figure}[ht!]
\centering
\includegraphics[width=500px]{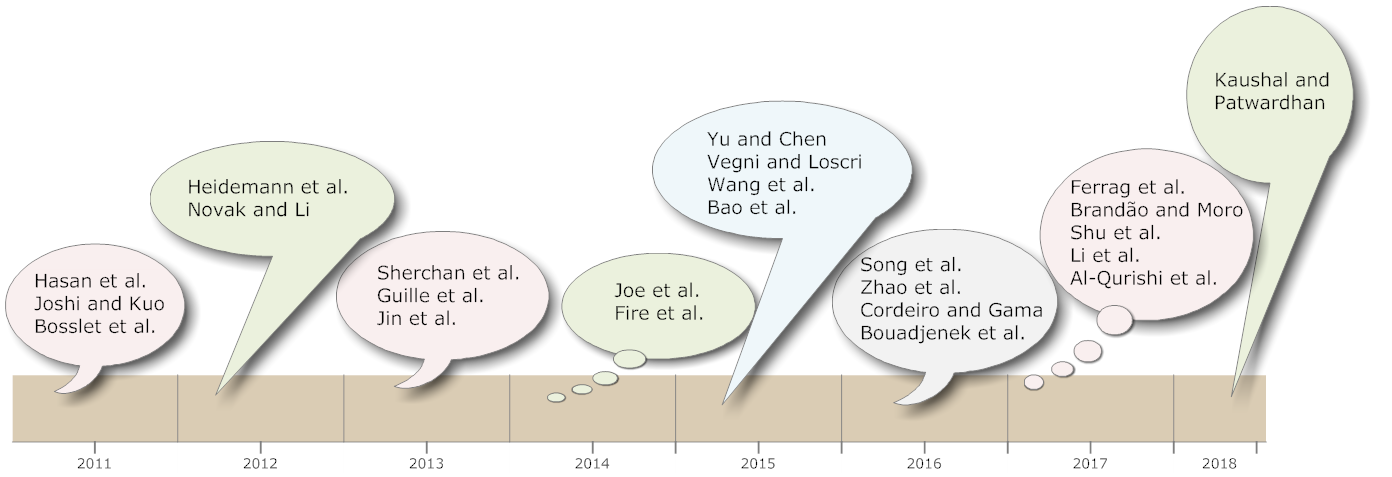}
\caption{A roadmap of existing surveys and their progressive comparison.}
\label{fig4}
\end{figure}
\section{Comparison with existing surveys}
\label{sect:Comparison}
POSNs aim at providing a reliable platform to make relationships among groups and communities. The emerging effect of POSNs also touches the business market in terms of advertising and dealing.

Many surveys have been conducted to evaluate the challenges and issues of POSNs as highlighted in Figure~\ref{fig4}. Some of the surveys fixate to find the trends and the technological enhancement of POSNs. The existing surveys address the problem with the social trends and their impact on the personal and the professional life of users involved in POSNs. Heidemann \emph{et al.} \cite{heidemann2012online} covers functionalities and characteristics of OSNs in their survey. Their survey emphasized emerging trends in OSNs and how advancement takes place with time. The role of OSNs in business perspective is also considered in their survey. Kaushal and Patwardhan \cite{kaushal2018emerging} focus on the personality trends of users in POSNs. The personality is an attribute or a characteristic of any user and it can be formed on the basis of behavioral, temperamental, emotional, and mental. The authors target the existing approaches and methods for predicting a user’s personality and their respective challenges. Their survey shows a tradeoff between the user of POSNs and personality and what linguistic features can be extracted from POSNs to analyze user’s profile.

Trust is an important entity for the OSNs. Sherchan \emph{et al.} \cite{sherchan2013survey} published a survey on social trust with OSNs. The author considers the information collection of trust, evaluation of the trust, and trust dissemination. Their survey highlights the buildup of social trust systems. Joshi and Kuo \cite{joshi2011security} presented a survey on the security and privacy issues in OSNs. The authors discussed how privacy is protected in the static and dynamic networks and what techniques exist to follow such a requirement. In the concern of security and privacy of OSNs, Joe and Ramakrishnan \cite{joe2014survey} presented a survey on various security issues in OSNs. Furthermore, Fire \emph{et al.}\cite{fire2014online} discussed the security and privacy risks in OSNs and existing solutions which are used for protection, security, and privacy of OSN users. The authors emphasized the threats of OSNs and the available commercial solutions. Novak and Li \cite{novak2012survey} presented the survey on the security and privacy issues in OSNs and also presented a review of user data protection mechanisms against malicious attacks. The anonymization and de-anonymization of OSN are also discussed in their survey. Ferrag \emph{et al.} \cite{ferrag2017privacy} gave a survey on the privacy-preserving schemes for ad-hoc social networks. The models including location privacy, identity privacy, anonymity, traceability, interest privacy, backward privacy, and content-oriented privacy are presented in their survey. In addition, the threats and vulnerabilities are addressed for the mobile social networks (MSNs) and vehicular social networks (VSNs). The challenges of privacy preservation and recommendations for further research are also given in their survey.

Brandão and Moro \cite{brandao2017social} focused on the social-professional networks. The authors presented the taxonomy for professional networks and their issues. The survey identified relationships among clustering, recommendation and ranking approaches of the social-professional networks.
Bouadjenek \emph{et al.} \cite{bouadjenek2016social} presented the survey on the social information retrieval approaches and platforms for OSNs. The survey included the review of important contributions and their analyses. Al-Qurishi \emph{et al.} \cite{al2017sybil} published a survey on the Sybil defense techniques in OSNs. The survey focused on the attacks that are possible in OSNs and their defensive schemes and methodologies for prevention. The survey included existing machine learning based solutions like supervised machine learning or unsupervised machine learning to detect Sybil attacks in OSNs. This survey classified all the related schemes based on the methods, datasets, and measurements. The classification of performance measurement and social network datasets of OSNs used in the literature are also given in their survey. A survey in event detection techniques of OSNs is given by Cordeiro and Gama \cite{cordeiro2016online}. The challenges faced by OSN in event detection are the main highlight of this survey.

Li \emph{et al.} \cite{li2017survey} provided state-of-the-art on link recommendation methods for OSNs. This survey mainly focused on the learning-based methods and proximity-based methods. The authors emphasized the theoretical foundations for link recommendation methods and their future directions. Al Hasan and Zaki \cite{al2011survey} gave a survey on the link prediction in OSNs. The authors analyzed existing link prediction models with their strengths and weaknesses in accurately identifying links in OSNs. In the era of link prediction in OSN, Wang \emph{et al.} \cite{wang2015link} presented the link prediction techniques and their problems. The link prediction can be obtained by mining and analyzing those helps to find missing links. Their survey mainly focused on the topology-based metrics and learning-based methods for link predictions. The applications and roadmaps are well addressed by the authors.

Bao \emph{et al.} \cite{bao2015recommendations} wrote a survey on the recommendations in location-based social networks. The survey focused on data sources and techniques that are used for recommendations. The comparative analysis of the recommender systems is analyzed in their survey. Yu and Chen \cite{yu2015survey} gave a survey on Point-of-interest (POI) recommendation in location-based social networks. Moreover, POI recommendation in location-based social networks is also discussed by Zhao \emph{et al.} \cite{zhao2016survey}.  The survey considers the aspects of POI recommendation, methodologies, and tasks for classification of taxonomies. The contributions and system features are highlighted for each aspect in their survey.

Song \emph{et al.} \cite{song2016trusting} investigated the cultural variations of different countries with respective of health information on the social networks. The authors mainly focused on the United States, Korea, and Hong Kong. The authors created a hypothesis on the basis of trust and use of experience-based knowledge shared on social Internet sites. Vegni and Loscri \cite{vegni2015survey} gave a survey on the main features of vehicular social networks and the major issues in bridging social networks to vehicular networking are discussed in their article.

Guille \emph{et al.} \cite{guille2013information} presented a survey on methods related to information diffusion analysis in OSNs. The authors discussed the challenges related to information diffusion in OSNs and also discussed the alternative approaches for a possible solution. Jin \emph{et al.} \cite{jin2013understanding} wrote a survey on the user behavior in OSNs. The survey discusses the interaction and connectivity of users and analysis of traffic activities. In order to further enhance the understanding of existing works, a comparison is presented in Table~\ref{Table5}.

\begin{landscape}
\begin{center}
\fontsize{8}{10}\selectfont
\setlength\LTleft{30pt}            
\setlength\LTright{0pt}
\begin{longtable}{@{\extracolsep{\fill}}*{8}{c}}
\caption{A comparison of the proposed survey and the existing surveys on OSNs/POSNs.}\label{Table5} \\
\hline
\multicolumn{1}{p{2.2cm}}{\centering \textbf{Survey}}
&\multicolumn{1}{p{2.0cm}}{\centering \textbf{Year}}
&\multicolumn{1}{p{1.5cm}}{\centering \textbf{Security} }
&\multicolumn{1}{p{1.5cm}}{\centering \textbf{Privacy}}
&\multicolumn{1}{p{1.5cm}}{\centering \textbf{Trust}}
&\multicolumn{1}{p{1.5cm}}{\centering \textbf{Anomaly detection}}
&\multicolumn{1}{p{1.5cm}}{\centering \textbf{OSNs/POSNs}}\\[6pt]\\
\hline \\

\endfirsthead

\multicolumn{8}{c}%
{{\bfseries \tablename\ \thetable{} -- continued from previous page}} \\
\hline
\multicolumn{1}{p{2.2cm}}{\centering \textbf{Survey}}
&\multicolumn{1}{p{2cm}}{\centering \textbf{Year}}
&\multicolumn{1}{p{1.5cm}}{\centering \textbf{Security} }
&\multicolumn{1}{p{1.5cm}}{\centering \textbf{Privacy}}
&\multicolumn{1}{p{1.5cm}}{\centering \textbf{Trust}}
&\multicolumn{1}{p{1.5cm}}{\centering \textbf{Anomaly detection}}
&\multicolumn{1}{p{1.5cm}}{\centering \textbf{OSNs/POSNs}}\\[6pt]\\
\hline\\\endhead

\hline \multicolumn{7}{l}{{Continued on next page}} \\

\endfoot

\endlastfoot

\multicolumn{1}{p{2.2cm}}{\centering  Proposed}
&\multicolumn{1}{p{2cm}}{\centering 2018}
&\multicolumn{1}{p{1.5cm}}{\centering \cmark}
&\multicolumn{1}{p{1.5cm}}{\centering \cmark}
&\multicolumn{1}{p{1.5cm}}{\centering  \cmark}
&\multicolumn{1}{p{1.5cm}}{\centering \cmark}
&\multicolumn{1}{p{1.5cm}}{\centering POSNs} \\[6pt]\\

\multicolumn{1}{p{2.2cm}}{\centering  Kaushal and Patwardhan \cite{kaushal2018emerging} }
&\multicolumn{1}{p{2cm}}{\centering 2018}
&\multicolumn{1}{p{1.5cm}}{\centering \xmark}
&\multicolumn{1}{p{1.5cm}}{\centering \xmark}
&\multicolumn{1}{p{1.5cm}}{\centering \xmark}
&\multicolumn{1}{p{1.5cm}}{\centering \xmark}
&\multicolumn{1}{p{1.5cm}}{\centering OSNs} \\[6pt]\\

\multicolumn{1}{p{2.2cm}}{\centering  Al-Qurishi \emph{et al.} \cite{al2017sybil} }
&\multicolumn{1}{p{2cm}}{\centering 2017}
&\multicolumn{1}{p{1.5cm}}{\centering \cmark}
&\multicolumn{1}{p{1.5cm}}{\centering \cmark}
&\multicolumn{1}{p{1.5cm}}{\centering \cmark}
&\multicolumn{1}{p{1.5cm}}{\centering \xmark}
&\multicolumn{1}{p{1.5cm}}{\centering OSNs} \\[6pt]\\

\multicolumn{1}{p{2.2cm}}{\centering  Li \emph{et al.} \cite{li2017survey} }
&\multicolumn{1}{p{2cm}}{\centering 2017}
&\multicolumn{1}{p{1.5cm}}{\centering \xmark}
&\multicolumn{1}{p{1.5cm}}{\centering \xmark}
&\multicolumn{1}{p{1.5cm}}{\centering \cmark}
&\multicolumn{1}{p{1.5cm}}{\centering \xmark}
&\multicolumn{1}{p{1.5cm}}{\centering OSNs} \\[6pt]\\

\multicolumn{1}{p{2.2cm}}{\centering  Brandão and Moro \cite{brandao2017social} }
&\multicolumn{1}{p{2cm}}{\centering 2017}
&\multicolumn{1}{p{1.5cm}}{\centering \xmark}
&\multicolumn{1}{p{1.5cm}}{\centering \xmark}
&\multicolumn{1}{p{1.5cm}}{\centering \xmark}
&\multicolumn{1}{p{1.5cm}}{\centering \xmark}
&\multicolumn{1}{p{1.5cm}}{\centering OSNs} \\[6pt]\\

\multicolumn{1}{p{2.2cm}}{\centering Ferrag \emph{et al.} \cite{ferrag2017privacy} }
&\multicolumn{1}{p{2cm}}{\centering 2017}
&\multicolumn{1}{p{1.5cm}}{\centering \cmark}
&\multicolumn{1}{p{1.5cm}}{\centering \cmark}
&\multicolumn{1}{p{1.5cm}}{\centering \cmark}
&\multicolumn{1}{p{1.5cm}}{\centering \cmark}
&\multicolumn{1}{p{1.5cm}}{\centering OSNs} \\[6pt]\\

\multicolumn{1}{p{2.2cm}}{\centering Bouadjenek \emph{et al.} \cite{bouadjenek2016social} }
&\multicolumn{1}{p{2cm}}{\centering 2016}
&\multicolumn{1}{p{1.5cm}}{\centering \xmark}
&\multicolumn{1}{p{1.5cm}}{\centering \cmark}
&\multicolumn{1}{p{1.5cm}}{\centering \cmark}
&\multicolumn{1}{p{1.5cm}}{\centering \xmark}
&\multicolumn{1}{p{1.5cm}}{\centering OSNs} \\[6pt]\\

\multicolumn{1}{p{2.2cm}}{\centering Cordeiro and Gama \cite{cordeiro2016online} }
&\multicolumn{1}{p{2cm}}{\centering 2016}
&\multicolumn{1}{p{1.5cm}}{\centering \xmark}
&\multicolumn{1}{p{1.5cm}}{\centering \xmark}
&\multicolumn{1}{p{1.5cm}}{\centering \xmark}
&\multicolumn{1}{p{1.5cm}}{\centering \xmark}
&\multicolumn{1}{p{1.5cm}}{\centering OSNs} \\[6pt]\\

\multicolumn{1}{p{2.2cm}}{\centering  Zhao \emph{et al.} \cite{zhao2016survey} }
&\multicolumn{1}{p{2cm}}{\centering 2016}
&\multicolumn{1}{p{1.5cm}}{\centering \xmark}
&\multicolumn{1}{p{1.5cm}}{\centering \xmark}
&\multicolumn{1}{p{1.5cm}}{\centering \xmark}
&\multicolumn{1}{p{1.5cm}}{\centering \xmark}
&\multicolumn{1}{p{1.5cm}}{\centering OSNs} \\[6pt]\\

\multicolumn{1}{p{2.2cm}}{\centering  Song \emph{et al.} \cite{song2016trusting} }
&\multicolumn{1}{p{2cm}}{\centering 2016}
&\multicolumn{1}{p{1.5cm}}{\centering \xmark}
&\multicolumn{1}{p{1.5cm}}{\centering \xmark}
&\multicolumn{1}{p{1.5cm}}{\centering \cmark}
&\multicolumn{1}{p{1.5cm}}{\centering \xmark}
&\multicolumn{1}{p{1.5cm}}{\centering OSNs} \\[6pt]\\

\multicolumn{1}{p{2.2cm}}{\centering  Bao \emph{et al.} \cite{bao2015recommendations} }
&\multicolumn{1}{p{2cm}}{\centering 2015}
&\multicolumn{1}{p{1.5cm}}{\centering \xmark}
&\multicolumn{1}{p{1.5cm}}{\centering \xmark}
&\multicolumn{1}{p{1.5cm}}{\centering \xmark}
&\multicolumn{1}{p{1.5cm}}{\centering \xmark}
&\multicolumn{1}{p{1.5cm}}{\centering OSNs} \\[6pt]\\

\multicolumn{1}{p{2.2cm}}{\centering  Wang \emph{et al.} \cite{wang2015link} }
&\multicolumn{1}{p{2cm}}{\centering 2015}
&\multicolumn{1}{p{1.5cm}}{\centering \xmark}
&\multicolumn{1}{p{1.5cm}}{\centering \xmark}
&\multicolumn{1}{p{1.5cm}}{\centering \xmark}
&\multicolumn{1}{p{1.5cm}}{\centering \xmark}
&\multicolumn{1}{p{1.5cm}}{\centering OSNs} \\[6pt]\\

\multicolumn{1}{p{2.2cm}}{\centering  Vegni and Loscri \cite{vegni2015survey} }
&\multicolumn{1}{p{2cm}}{\centering 2015}
&\multicolumn{1}{p{1.5cm}}{\centering \cmark}
&\multicolumn{1}{p{1.5cm}}{\centering \cmark}
&\multicolumn{1}{p{1.5cm}}{\centering \cmark}
&\multicolumn{1}{p{1.5cm}}{\centering \xmark}
&\multicolumn{1}{p{1.5cm}}{\centering OSNs} \\[6pt]\\

\multicolumn{1}{p{2.2cm}}{\centering  Yu and Chen \cite{yu2015survey} }
&\multicolumn{1}{p{2cm}}{\centering 2015}
&\multicolumn{1}{p{1.5cm}}{\centering \xmark}
&\multicolumn{1}{p{1.5cm}}{\centering \xmark}
&\multicolumn{1}{p{1.5cm}}{\centering \cmark}
&\multicolumn{1}{p{1.5cm}}{\centering \xmark}
&\multicolumn{1}{p{1.5cm}}{\centering OSNs} \\[6pt]\\

\multicolumn{1}{p{2.2cm}}{\centering Fire \emph{et al.}\cite{fire2014online} }
&\multicolumn{1}{p{2cm}}{\centering 2014}
&\multicolumn{1}{p{1.5cm}}{\centering \cmark}
&\multicolumn{1}{p{1.5cm}}{\centering \cmark}
&\multicolumn{1}{p{1.5cm}}{\centering \cmark}
&\multicolumn{1}{p{1.5cm}}{\centering \cmark}
&\multicolumn{1}{p{1.5cm}}{\centering OSNs} \\[6pt]\\

\multicolumn{1}{p{2.2cm}}{\centering Joe and Ramakrishnan \cite{joe2014survey} }
&\multicolumn{1}{p{2cm}}{\centering 2014}
&\multicolumn{1}{p{1.5cm}}{\centering \cmark}
&\multicolumn{1}{p{1.5cm}}{\centering \cmark}
&\multicolumn{1}{p{1.5cm}}{\centering \xmark}
&\multicolumn{1}{p{1.5cm}}{\centering \cmark}
&\multicolumn{1}{p{1.5cm}}{\centering OSNs} \\[6pt]\\

\multicolumn{1}{p{2.2cm}}{\centering Jin \emph{et al.} \cite{jin2013understanding} }
&\multicolumn{1}{p{2cm}}{\centering 2013}
&\multicolumn{1}{p{1.5cm}}{\centering \cmark}
&\multicolumn{1}{p{1.5cm}}{\centering \cmark}
&\multicolumn{1}{p{1.5cm}}{\centering \cmark}
&\multicolumn{1}{p{1.5cm}}{\centering \cmark}
&\multicolumn{1}{p{1.5cm}}{\centering OSNs} \\[6pt]\\

\multicolumn{1}{p{2.2cm}}{\centering Guille \emph{et al.} \cite{guille2013information} }
&\multicolumn{1}{p{2cm}}{\centering 2013}
&\multicolumn{1}{p{1.5cm}}{\centering \xmark}
&\multicolumn{1}{p{1.5cm}}{\centering \xmark}
&\multicolumn{1}{p{1.5cm}}{\centering \xmark}
&\multicolumn{1}{p{1.5cm}}{\centering \cmark}
&\multicolumn{1}{p{1.5cm}}{\centering OSNs} \\[6pt]\\

\multicolumn{1}{p{2.2cm}}{\centering Sherchan \emph{et al.} \cite{sherchan2013survey} }
&\multicolumn{1}{p{2cm}}{\centering 2013}
&\multicolumn{1}{p{1.5cm}}{\centering \cmark}
&\multicolumn{1}{p{1.5cm}}{\centering \cmark}
&\multicolumn{1}{p{1.5cm}}{\centering \cmark}
&\multicolumn{1}{p{1.5cm}}{\centering \xmark}
&\multicolumn{1}{p{1.5cm}}{\centering OSNs} \\[6pt]\\

\multicolumn{1}{p{2.2cm}}{\centering Novak and Li \cite{novak2012survey} }
&\multicolumn{1}{p{2cm}}{\centering 2013}
&\multicolumn{1}{p{1.5cm}}{\centering \cmark}
&\multicolumn{1}{p{1.5cm}}{\centering \cmark}
&\multicolumn{1}{p{1.5cm}}{\centering \cmark}
&\multicolumn{1}{p{1.5cm}}{\centering \cmark}
&\multicolumn{1}{p{1.5cm}}{\centering OSNs} \\[6pt]\\

\multicolumn{1}{p{2.2cm}}{\centering Heidemann \emph{et al.} \cite{heidemann2012online} }
&\multicolumn{1}{p{2cm}}{\centering 2012}
&\multicolumn{1}{p{1.5cm}}{\centering \cmark}
&\multicolumn{1}{p{1.5cm}}{\centering \cmark}
&\multicolumn{1}{p{1.5cm}}{\centering \cmark}
&\multicolumn{1}{p{1.5cm}}{\centering \xmark}
&\multicolumn{1}{p{1.5cm}}{\centering OSNs} \\[6pt]\\

\multicolumn{1}{p{2.2cm}}{\centering Bosslet \emph{et al.} \cite{bosslet2011patient} }
&\multicolumn{1}{p{2cm}}{\centering 2011}
&\multicolumn{1}{p{1.5cm}}{\centering \xmark}
&\multicolumn{1}{p{1.5cm}}{\centering \xmark}
&\multicolumn{1}{p{1.5cm}}{\centering \xmark}
&\multicolumn{1}{p{1.5cm}}{\centering \xmark}
&\multicolumn{1}{p{1.5cm}}{\centering OSNs} \\[6pt]\\

\multicolumn{1}{p{2.2cm}}{\centering Joshi and Kuo \cite{joshi2011security} }
&\multicolumn{1}{p{2cm}}{\centering 2011}
&\multicolumn{1}{p{1.5cm}}{\centering \cmark}
&\multicolumn{1}{p{1.5cm}}{\centering \cmark}
&\multicolumn{1}{p{1.5cm}}{\centering \cmark}
&\multicolumn{1}{p{1.5cm}}{\centering \cmark}
&\multicolumn{1}{p{1.5cm}}{\centering OSNs} \\[6pt]\\

\multicolumn{1}{p{2.2cm}}{\centering Al Hasan and Zaki \cite{al2011survey} }
&\multicolumn{1}{p{2cm}}{\centering 2011}
&\multicolumn{1}{p{1.5cm}}{\centering \xmark}
&\multicolumn{1}{p{1.5cm}}{\centering \xmark}
&\multicolumn{1}{p{1.5cm}}{\centering \xmark}
&\multicolumn{1}{p{1.5cm}}{\centering \xmark}
&\multicolumn{1}{p{1.5cm}}{\centering OSNs} \\[6pt]\\

\hline

\end{longtable}
\end{center}

\begin{center}
\fontsize{7}{10}\selectfont
\setlength\LTleft{30pt}            
\setlength\LTright{0pt}
\begin{longtable}{@{\extracolsep{\fill}}*{10}{c}}
\caption{State-of-the-art solutions for security enhancement of POSNs.}\label{Table2} \\
\hline
\multicolumn{1}{p{2.2cm}}{\centering \textbf{Approach}}
&\multicolumn{1}{p{2.2cm}}{\centering \textbf{Author} }
&\multicolumn{1}{p{2.2cm}}{\centering \textbf{Ideology}}
&\multicolumn{1}{p{2cm}}{\centering \textbf{Application} }
&\multicolumn{1}{p{2.2cm}}{\centering \textbf{Parameter}}
&\multicolumn{1}{p{1.2cm}}{\centering \textbf{Security}}
&\multicolumn{1}{p{1.0cm}}{\centering \textbf{Privacy}}
&\multicolumn{1}{p{1.0cm}}{\centering \textbf{Authentication}}
&\multicolumn{1}{p{1.0cm}}{\centering \textbf{Encryption}}
&\multicolumn{1}{p{1.0cm}}{\centering \textbf{Data Analytic}}\\[6pt]\\
\hline \\

\endfirsthead

\multicolumn{10}{c}%
{{\bfseries \tablename\ \thetable{} -- continued from previous page}} \\
\hline
\multicolumn{1}{p{2.2cm}}{\centering \textbf{Approach}}
&\multicolumn{1}{p{2.2cm}}{\centering \textbf{Author} }
&\multicolumn{1}{p{2.2cm}}{\centering \textbf{Ideology}}
&\multicolumn{1}{p{1.5cm}}{\centering \textbf{Application} }
&\multicolumn{1}{p{2.2cm}}{\centering \textbf{Parameter}}
&\multicolumn{1}{p{1.0cm}}{\centering \textbf{Security}}
&\multicolumn{1}{p{1.0cm}}{\centering \textbf{Privacy}}
&\multicolumn{1}{p{1.2cm}}{\centering \textbf{Authentication}}
&\multicolumn{1}{p{1cm}}{\centering \textbf{Encryption}}
&\multicolumn{1}{p{1.0cm}}{\centering \textbf{Data Analytic}}\\[6pt]\\
\hline\\\endhead

\hline \multicolumn{10}{l}{{Continued on next page}} \\

\endfoot

\endlastfoot

\multicolumn{1}{p{2.2cm}}{\centering  CenLocShare }
&\multicolumn{1}{p{2.2cm}}{\centering [Xiao \emph{et al.} 2017]\cite{xiao2017cenlocshare}}
&\multicolumn{1}{p{2.2cm}}{\centering Centralized privacy-preserving location-sharing system}
&\multicolumn{1}{p{1.5cm}}{\centering mOSNs}
&\multicolumn{1}{p{2.2cm}}{\centering Time of location query, Storage space, client interfaces}
&\multicolumn{1}{p{1.0cm}}{\centering \cmark}
&\multicolumn{1}{p{1.0cm}}{\centering \cmark}
&\multicolumn{1}{p{1.2cm}}{\centering \xmark}
&\multicolumn{1}{p{1cm}}{\centering \cmark}
&\multicolumn{1}{p{1.0cm}}{\centering On-site} \\[6pt]\\

\multicolumn{1}{p{2.2cm}}{\centering  UDPLS }
&\multicolumn{1}{p{2.2cm}}{\centering [Suno \emph{et al.}2017]\cite{sun2017user}}
&\multicolumn{1}{p{2.2cm}}{\centering Preserve user's location privacy and network privacy on location server, and preserve user's location privacy on social network server}
&\multicolumn{1}{p{1.5cm}}{\centering mOSNs}
&\multicolumn{1}{p{2.2cm}}{\centering Computation time,query time}
&\multicolumn{1}{p{1.0cm}}{\centering \cmark}
&\multicolumn{1}{p{1.0cm}}{\centering \cmark}
&\multicolumn{1}{p{1.2cm}}{\centering \cmark}
&\multicolumn{1}{p{1cm}}{\centering \cmark}
&\multicolumn{1}{p{1.0cm}}{\centering On-site} \\[6pt]\\

\multicolumn{1}{p{2.2cm}}{\centering Safebook }
&\multicolumn{1}{p{2.2cm}}{\centering [Antonio Cutillo and Molva 2009 ]\cite{cutillo2009safebook}}
&\multicolumn{1}{p{2.2cm}}{\centering Cooperation among a number of independent parties}
&\multicolumn{1}{p{1.5cm}}{\centering OSNs}
&\multicolumn{1}{p{2.2cm}}{\centering Entrypoint registration, data lookup and retrieval}
&\multicolumn{1}{p{1.0cm}}{\centering \cmark}
&\multicolumn{1}{p{1.0cm}}{\centering \cmark}
&\multicolumn{1}{p{1.2cm}}{\centering \cmark}
&\multicolumn{1}{p{1cm}}{\centering \cmark}
&\multicolumn{1}{p{1.0cm}}{\centering On-site} \\[6pt]\\

\multicolumn{1}{p{2.2cm}}{\centering Secure friend discovery }
&\multicolumn{1}{p{2.2cm}}{\centering [Dong \emph{et al.}2011]\cite{dong2011secure}}
&\multicolumn{1}{p{2.2cm}}{\centering Compute social proximity between two users to discover potential friends}
&\multicolumn{1}{p{1.5cm}}{\centering mOSNs}
&\multicolumn{1}{p{2.2cm}}{\centering CDF of cosine similarity of social coordinates}
&\multicolumn{1}{p{1.0cm}}{\centering \cmark}
&\multicolumn{1}{p{1.0cm}}{\centering \cmark}
&\multicolumn{1}{p{1.2cm}}{\centering \cmark}
&\multicolumn{1}{p{1cm}}{\centering \cmark}
&\multicolumn{1}{p{1.0cm}}{\centering On-site} \\[6pt]\\

\multicolumn{1}{p{2.2cm}}{\centering SuperNova }
&\multicolumn{1}{p{2.2cm}}{\centering [Sharma and Datta 2012]\cite{sharma2012supernova}}
&\multicolumn{1}{p{2.2cm}}{\centering Heuristics approach on the  end user resources and their behaviors}
&\multicolumn{1}{p{1.5cm}}{\centering DOSNs}
&\multicolumn{1}{p{2.2cm}}{\centering Cumulative availability, system performance}
&\multicolumn{1}{p{1.0cm}}{\centering \cmark}
&\multicolumn{1}{p{1.0cm}}{\centering -}
&\multicolumn{1}{p{1.2cm}}{\centering \xmark}
&\multicolumn{1}{p{1cm}}{\centering \xmark}
&\multicolumn{1}{p{1.0cm}}{\centering -} \\[6pt]\\

\multicolumn{1}{p{2.2cm}}{\centering DECENT }
&\multicolumn{1}{p{2.2cm}}{\centering [Jahid \emph{et al.}2012]\cite{jahid2012decent}}
&\multicolumn{1}{p{2.2cm}}{\centering Distributed hash table to store user data, and features cryptographic protections}
&\multicolumn{1}{p{1.5cm}}{\centering OSNs}
&\multicolumn{1}{p{2.2cm}}{\centering Average time to view a newsfeed}
&\multicolumn{1}{p{1.0cm}}{\centering \cmark}
&\multicolumn{1}{p{1.0cm}}{\centering \cmark}
&\multicolumn{1}{p{1.2cm}}{\centering \cmark}
&\multicolumn{1}{p{1cm}}{\centering \cmark}
&\multicolumn{1}{p{1.0cm}}{\centering Off-site} \\[6pt]\\

\multicolumn{1}{p{2.2cm}}{\centering Protection of PSN  }
&\multicolumn{1}{p{2.2cm}}{\centering [Yan and Wang 2017] \cite{yan2017protect}}
&\multicolumn{1}{p{2.2cm}}{\centering Two-dimensional trust levels}
&\multicolumn{1}{p{1.5cm}}{\centering POSNs}
&\multicolumn{1}{p{2.2cm}}{\centering Computational complexity, secret key generation time}
&\multicolumn{1}{p{1.0cm}}{\centering \cmark}
&\multicolumn{1}{p{1.0cm}}{\centering \cmark}
&\multicolumn{1}{p{1.2cm}}{\centering \xmark}
&\multicolumn{1}{p{1cm}}{\centering \cmark}
&\multicolumn{1}{p{1.0cm}}{\centering -} \\[6pt]\\

\multicolumn{1}{p{2.2cm}}{\centering ProGuard  }
&\multicolumn{1}{p{2.2cm}}{\centering [Zhou \emph{et al.}2017]  \cite{zhou2017proguard}}
&\multicolumn{1}{p{2.2cm}}{\centering Detecting malicious accounts}
&\multicolumn{1}{p{1.5cm}}{\centering OSNs}
&\multicolumn{1}{p{2.2cm}}{\centering Detection rate, false positive rate, total amount of expenditure}
&\multicolumn{1}{p{1.0cm}}{\centering -}
&\multicolumn{1}{p{1.0cm}}{\centering -}
&\multicolumn{1}{p{1.2cm}}{\centering \xmark}
&\multicolumn{1}{p{1cm}}{\centering \xmark}
&\multicolumn{1}{p{1.0cm}}{\centering Off-site} \\[6pt]\\

\multicolumn{1}{p{2.2cm}}{\centering Private data publication  }
&\multicolumn{1}{p{2.2cm}}{\centering [Zheng \emph{et al.}2018] \cite{zheng2018fair}}
&\multicolumn{1}{p{2.2cm}}{\centering Heterogeneous privacy preferences and the correlations among participants}
&\multicolumn{1}{p{1.5cm}}{\centering OSNs}
&\multicolumn{1}{p{2.2cm}}{\centering Performance for Heterogeneous Users, ratio of the successfully served users}
&\multicolumn{1}{p{1.0cm}}{\centering \xmark}
&\multicolumn{1}{p{1.0cm}}{\centering \cmark}
&\multicolumn{1}{p{1.2cm}}{\centering \xmark}
&\multicolumn{1}{p{1cm}}{\centering \xmark}
&\multicolumn{1}{p{1.0cm}}{\centering Off-site} \\[6pt]\\

\multicolumn{1}{p{2.2cm}}{\centering NHAD  }
&\multicolumn{1}{p{2.2cm}}{\centering [Sharma  \emph{et al.}2018]\cite{8322278}}
&\multicolumn{1}{p{2.2cm}}{\centering Used paradigms-missing links, reputation gain, significant difference, trust properties, and trust score}
&\multicolumn{1}{p{1.5cm}}{\centering OSNs}
&\multicolumn{1}{p{2.2cm}}{\centering Detection rate, false postive rate, Accuracy, F-score, precision}
&\multicolumn{1}{p{1.0cm}}{\centering \xmark}
&\multicolumn{1}{p{1.0cm}}{\centering \xmark}
&\multicolumn{1}{p{1.2cm}}{\centering \xmark}
&\multicolumn{1}{p{1cm}}{\centering \xmark}
&\multicolumn{1}{p{1.0cm}}{\centering On-site} \\[6pt]\\

\multicolumn{1}{p{2.2cm}}{\centering SybilTrap  }
&\multicolumn{1}{p{2.2cm}}{\centering [Al-Qurishi \emph{et al.}2018]\cite{al2018sybiltrap}}
&\multicolumn{1}{p{2.2cm}}{\centering Graph based supervised learning technique}
&\multicolumn{1}{p{1.5cm}}{\centering OSNs}
&\multicolumn{1}{p{2.2cm}}{\centering Receiver operating characteristic,cumulative distribution function (CDF) of each feature}
&\multicolumn{1}{p{1.0cm}}{\centering \xmark}
&\multicolumn{1}{p{1.0cm}}{\centering \xmark}
&\multicolumn{1}{p{1.2cm}}{\centering \xmark}
&\multicolumn{1}{p{1cm}}{\centering \xmark}
&\multicolumn{1}{p{1.0cm}}{\centering On-site} \\[6pt]\\

\multicolumn{1}{p{2.2cm}}{\centering lurkers detection  }
&\multicolumn{1}{p{2.2cm}}{\centering [Amatoi \emph{et al.}2018]\cite{amato2018centrality}}
&\multicolumn{1}{p{2.2cm}}{\centering Based on hypergraphs}
&\multicolumn{1}{p{1.5cm}}{\centering Heterogeneous OSNs}
&\multicolumn{1}{p{2.2cm}}{\centering Loading times, running times}
&\multicolumn{1}{p{1.0cm}}{\centering \xmark}
&\multicolumn{1}{p{1.0cm}}{\centering \xmark}
&\multicolumn{1}{p{1.2cm}}{\centering \xmark}
&\multicolumn{1}{p{1cm}}{\centering \xmark}
&\multicolumn{1}{p{1.0cm}}{\centering Off-site} \\[6pt]\\

\multicolumn{1}{p{2.2cm}}{\centering Automatic control/block over user comments  }
&\multicolumn{1}{p{2.2cm}}{\centering [Godse \emph{et al.}2018]\cite{godse2018automatic}}
&\multicolumn{1}{p{2.2cm}}{\centering User control system (UCS)}
&\multicolumn{1}{p{1.5cm}}{\centering OSNs}
&\multicolumn{1}{p{2.2cm}}{\centering Policy, database}
&\multicolumn{1}{p{1.0cm}}{\centering \cmark}
&\multicolumn{1}{p{1.0cm}}{\centering \cmark}
&\multicolumn{1}{p{1.2cm}}{\centering \xmark}
&\multicolumn{1}{p{1cm}}{\centering \xmark}
&\multicolumn{1}{p{1.0cm}}{\centering -} \\[6pt]\\

\multicolumn{1}{p{2.2cm}}{\centering De-Anonymizing framework }
&\multicolumn{1}{p{2.2cm}}{\centering [Su \emph{et al.}2017]\cite{su2017anonymizing}}
&\multicolumn{1}{p{2.2cm}}{\centering Browsing behaviour}
&\multicolumn{1}{p{1.5cm}}{\centering OSNs}
&\multicolumn{1}{p{2.2cm}}{\centering De-Anonymizing accuracy}
&\multicolumn{1}{p{1.0cm}}{\centering -}
&\multicolumn{1}{p{1.0cm}}{\centering -}
&\multicolumn{1}{p{1.2cm}}{\centering \xmark}
&\multicolumn{1}{p{1cm}}{\centering \xmark}
&\multicolumn{1}{p{1.0cm}}{\centering On-site} \\[6pt]\\

\multicolumn{1}{p{2.2cm}}{\centering Online social network for emergency management }
&\multicolumn{1}{p{2.2cm}}{\centering [Roxanne Hiltz and Turoff 2009 ]\cite{white2009online}}
&\multicolumn{1}{p{2.2cm}}{\centering Establishment of global relationships in case of  emergency}
&\multicolumn{1}{p{1.5cm}}{\centering OSNs}
&\multicolumn{1}{p{2.2cm}}{\centering Uses of SNSs for emergency management}
&\multicolumn{1}{p{1.0cm}}{\centering \xmark}
&\multicolumn{1}{p{1.0cm}}{\centering \xmark}
&\multicolumn{1}{p{1.2cm}}{\centering \xmark}
&\multicolumn{1}{p{1cm}}{\centering \xmark}
&\multicolumn{1}{p{1.0cm}}{\centering -} \\[6pt]\\

\hline
\end{longtable}
\end{center}
\end{landscape}

\begin{figure}[ht!]
\centering
\includegraphics[width=400px]{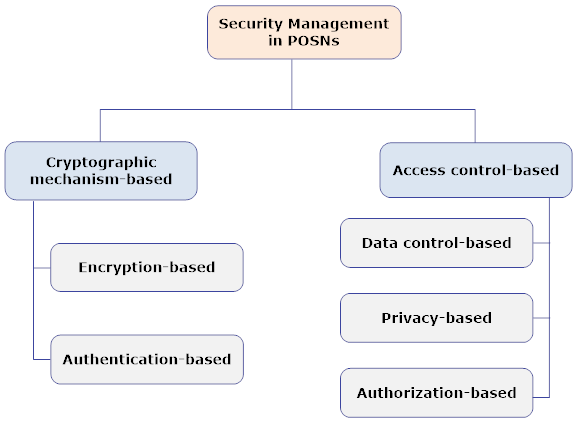}
\caption{A taxonomy of security solutions for POSNs.}
\label{fig_security}
\end{figure}
\section{Security in POSNs}
\label{sect:Taxonomy_security}
The users of OSNs, such as Facebook, Twitter, and Google+ are increasing exponentially and so is the possibility of multiple communities in futuristic POSNs. The security of this critical information is required for smooth transmissions in POSNs. The security should be integrated into the user’s information and communications. In this survey, the security management is classified into two parts, as shown in Figure~\ref{fig_security}:
\subsection{Cryptographic mechanism-based security}
Social media technology mainly uses cryptographic security techniques. Cryptographic techniques are suitable for groups with dynamic memberships~\cite{jahid2012decent}~\cite{hussain2017pbf}. The group is any community or any cluster which shows same properties. The social media problems of security, privacy and anti-piracy can be overcome through cryptographic techniques like authentication, encryption etc.
\subsubsection{Encryption-based security}
The encryption-based techniques encrypt the location information or all the information of a user. The communication security can be achieved through peer to peer encryption or peer to multi-peer encryption. There are many existing techniques like attribute-based encryption, proxy-based encryption schemes to protect accidental or intentional information leak in POSNs~\cite{zhou2017security}~\cite{qiu2018proactive}.
\subsubsection{Authentication-based security}
The authentication-based security is achieved by applying authentication protocols among users while sharing their personal information. The authentication is performed mutually either in an individual mode or in a group mode. The authentication schemes provide secure communication and a user-specific data accessibility~\cite{ma2014security}~\cite{wu2012robust}. The message authentication maintains the integrity in the context of POSNs.

\subsection{Access control-based security}
The access control based techniques are used for controlling the shared information. The access control provides restriction over unwanted information sharing in the public domain over the social media~\cite{cheng2012relationship}~\cite{tourani2017security}. The access control can be applied by a user or by a controller, or both in certain situations. The access control based security is further categorized with the attributes as explained below:

\subsubsection{Data control-based}
Data control refers to the amount of information that is shared among user to maintain anonymity~\cite{bahri2018enhanced}. The data is shared among many communities over different social media. The data leakage leads to the breach of the privacy and the integrity. The access control helps to maintain the integrity by providing limited access to unauthorized users~\cite{masoumzadeh2018security}.
\subsubsection{Privacy-based}
The distributed information over social site leads to privacy concerns and requires insights into security problems~\cite{sharma2017cooperative}~\cite{zhao2017privacy}. The privacy of users provides integrity in the interpersonal relationships and flexibility in POSNs. The privacy based techniques help to preserve the security of private information and prevent data from leakage.
\subsubsection{Authorization-based}
Authorization is a process of leveraging accessing information by authenticating legitimate users of POSNs. The services are accessible within their role over POSNs. The authorization helps to block the unwanted users from connecting communities or trying to build a relationship with other users~\cite{wang2013system}~\cite{tourani2017security}.

A detailed comparison of existing solutions on the security of POSNs is presented in Table~\ref{Table2}.

\begin{figure}[ht!]
\centering
\includegraphics[width=430px]{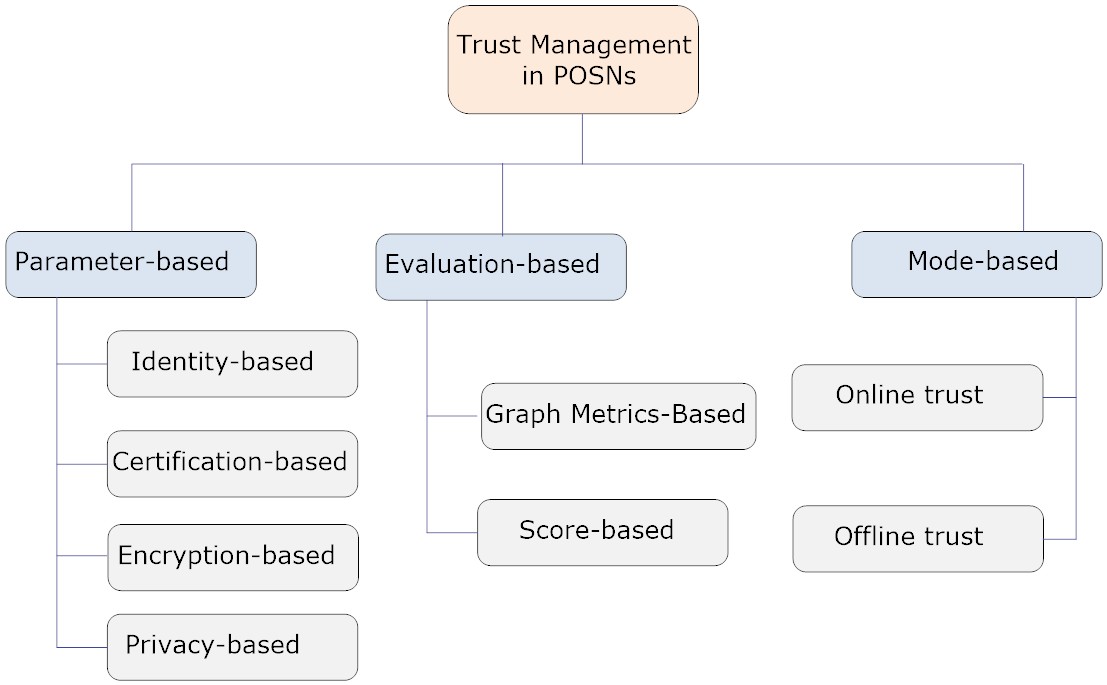}
\caption{A taxonomy of trust management approaches in POSNs.}
\label{tax_trust}
\end{figure}

\section{Trust Management in POSNs}
\label{sect:Taxonomy_trust}
Trust has been an important concern for POSNs. POSNs incorporate trust models and algorithms to enhance service qualities, user experiences, and reliability. Trust mechanism of these networks relies on several paradigms like properties, attributes, modes etc. The trust taxonomy, as shown in Figure~\ref{tax_trust}, represents the classification of trust management solutions in the POSNs~\cite{sharma2017computational}~\cite{8322278}.

\subsection{Parameter-based trust management}
The trust management relies on several cryptographic techniques which are used to protect information confidentiality and integrity of POSNs. The Trust is achieved by applying trust models and algorithms with a different set of parameters~\cite{sun2007distributed}~\cite{hsu2007knowledge}. The parameters are the attributes which are associated with the properties like privacy, identity etc. The parameter based trust management can be further classified into following subcategories.

\subsubsection{Identity-based}
The authorization and accounting procedures of the user over POSNs help to construct an Identity-based trust management which supports the transparency of communications and facilitates the agents to establish their reputation~\cite{premarathne2017cloud}~\cite{burmann2017concept}. The original Identity of the user is hidden and pseudo-identities are used to enhance the privacy.
\subsubsection{Certification-based}
The certificates and trusted certification authorities help to build a trust in POSNs by controlling the accessibility of users to services. The certification authorities help to provide central services for automatic public query and for validity checks on self-issued certificates~\cite{ardagna2018certification}~\cite{omar2012certification}.

\subsubsection{Encryption-based}
The encryption is used to hide the plain text and formulate into a new ciphertext. POSNs contain lots of private conversations and critical information of users~\cite{lotspiech2004anonymous}~\cite{shang2017breaking}. Encryption schemes-Role Based encryption, Homomorphic encryptions, Identity-based encryption, pseudonym-based encryption, attribute-based encryption are used with the POSNs to provide trust for users~\cite{tseng2017homomorphic}.

\subsubsection{Privacy-based}
Trust establishment is used as a measure of the perception of security in POSNs. Privacy preservation plays a significant role in trust management. The privacy access controls are established to control the users' information in public domains~\cite{premarathne2017cloud}~\cite{ruotsalainen2017privacy}. These privacy access solutions provide controls based on the data placed on profile pages such as name, pictures, etc.

\subsection{Evaluation-based trust management}
The interactions among the users rely on trust between the social networks and participants of POSNs. The reputation is usually built on feedback from those who have direct interactions. The reputation-based trust establishment is commonly used mechanism for reliability in POSNs~\cite{yan2017flexible}~\cite{guo2017mutual}~\cite{shen2017trustworthiness}. The trust evolution is relied on following two methods:

\subsubsection{Graph metrics-based}
In the graph metrics-based trust evaluations, the social networks consider a large graph with many users connected immediately or explicitly to each other. In the graph metrics, the trust value is calculated through a trusted graph~\cite{de2018discovering}. The trust establishment will help enhance the user's experience and the service quality.

\subsubsection{Score-based trust}
In the score based trust evaluations, the score is calculated for reputation building among users. The rating score is utilized by probability based models that are used for trust establishment~\cite{jabeen2018anonymity}. The quantitative trust is helpful to define the trust level over POSNs.

\subsection{Mode-based trust management}
In the mode-based trust management, the trust establishment is connected to the factor of trust which characterizes the information as well as the users in the online or offline mode of connections~\cite{he2016understanding}.

\subsubsection{Offline trust}
In the offline mode, the trust establishment is performed at the offsite over large data sets which are observed over a long duration of time. Offline trust helps to secure those systems which have a periodic break between their transactions, such as POSN-banking~\cite{8322278}~\cite{sharma2017isma}.

\subsubsection{Online trust}
In the online mode, the trust is established on-demand based on the online behaviors and the profiling outcomes such as the adoption of apps, downloads of content, voting, and re-sharing of contents. The online mode contains dynamic adoption of the respective commands used by the users for their accounts or data~\cite{8322278}~\cite{sharma2017isma}.

A detailed comparison of existing solutions on the trust management in POSNs is presented in Table~\ref{Table3}.
\begin{landscape}
\begin{center}
\fontsize{7}{10}\selectfont
\setlength\LTleft{30pt}            
\setlength\LTright{0pt}
\begin{longtable}{@{\extracolsep{\fill}}*{9}{c}}
\caption{State-of-the-art solutions for trust management in POSNs.}\label{Table3} \\
\hline
\multicolumn{1}{p{2.2cm}}{\centering \textbf{Approach}}
&\multicolumn{1}{p{2.2cm}}{\centering \textbf{Author} }
&\multicolumn{1}{p{2.2cm}}{\centering \textbf{Ideology}}
&\multicolumn{1}{p{2cm}}{\centering \textbf{Application} }
&\multicolumn{1}{p{2.2cm}}{\centering \textbf{Parameter}}
&\multicolumn{1}{p{1.2cm}}{\centering \textbf{Mechanism}}
&\multicolumn{1}{p{1.0cm}}{\centering \textbf{Anomaly Detection}}
&\multicolumn{1}{p{1.0cm}}{\centering \textbf{Visualization}}
&\multicolumn{1}{p{1.0cm}}{\centering \textbf{Reputation-Formation}}\\[6pt]\\
\hline \\

\endfirsthead

\multicolumn{9}{c}%
{{\bfseries \tablename\ \thetable{} -- continued from previous page}} \\
\hline
\multicolumn{1}{p{2.2cm}}{\centering \textbf{Approach}}
&\multicolumn{1}{p{2.2cm}}{\centering \textbf{Author} }
&\multicolumn{1}{p{2.2cm}}{\centering \textbf{Ideology}}
&\multicolumn{1}{p{2cm}}{\centering \textbf{Application} }
&\multicolumn{1}{p{2.2cm}}{\centering \textbf{Parameter}}
&\multicolumn{1}{p{1.2cm}}{\centering \textbf{Mechanism}}
&\multicolumn{1}{p{1.0cm}}{\centering \textbf{Anomaly Detection}}
&\multicolumn{1}{p{1.0cm}}{\centering \textbf{Visualization}}
&\multicolumn{1}{p{1.0cm}}{\centering \textbf{Reputation-Formation}}\\[6pt]\\
\hline\\\endhead

\hline \multicolumn{9}{l}{{Continued on next page}} \\

\endfoot

\endlastfoot

\multicolumn{1}{p{2.2cm}}{\centering  User-domain based trusted acquaintance chain discovery algorithm }
&\multicolumn{1}{p{2.2cm}}{\centering [Jiang \emph{et al.} 2014] \cite{jiang2014generating}}
&\multicolumn{1}{p{2.2cm}}{\centering Trusted graph with the adjustable width breadth-first search algorithms}
&\multicolumn{1}{p{1.5cm}}{\centering OSNs}
&\multicolumn{1}{p{2.2cm}}{\centering Trust conflict, quality of nodes}
&\multicolumn{1}{p{1.0cm}}{\centering Hybrid}
&\multicolumn{1}{p{1.0cm}}{\centering \xmark}
&\multicolumn{1}{p{1.2cm}}{\centering \xmark}
&\multicolumn{1}{p{1cm}}{\centering \cmark}\\\\

\multicolumn{1}{p{2.2cm}}{\centering  Probabilistic recommendation model}
&\multicolumn{1}{p{2.2cm}}{\centering [Wang \emph{et al.}2015]\cite{wang2015trust}}
&\multicolumn{1}{p{2.2cm}}{\centering Combining recommendation attributes with inherent similarity}
&\multicolumn{1}{p{1.5cm}}{\centering OSNs}
&\multicolumn{1}{p{2.2cm}}{\centering Precision, recall, transition probability influence factors}
&\multicolumn{1}{p{1.0cm}}{\centering -}
&\multicolumn{1}{p{1.0cm}}{\centering \xmark}
&\multicolumn{1}{p{1.2cm}}{\centering \xmark}
&\multicolumn{1}{p{1cm}}{\centering \cmark}\\\\

\multicolumn{1}{p{2.2cm}}{\centering  Access control scheme}
&\multicolumn{1}{p{2.2cm}}{\centering [Pang and Zhang 2015]\cite{pang2015new}}
&\multicolumn{1}{p{2.2cm}}{\centering Hybrid logic for formulating access control policies}
&\multicolumn{1}{p{1.5cm}}{\centering OSNs}
&\multicolumn{1}{p{2.2cm}}{\centering Information reliability, relationship hierarchy}
&\multicolumn{1}{p{1.0cm}}{\centering Distributed}
&\multicolumn{1}{p{1.0cm}}{\centering \xmark}
&\multicolumn{1}{p{1.2cm}}{\centering \cmark}
&\multicolumn{1}{p{1cm}}{\centering \cmark}\\\\

\multicolumn{1}{p{2.2cm}}{\centering  Consumer perception of knowledge-sharing model}
&\multicolumn{1}{p{2.2cm}}{\centering [Bilgihan \emph{et al.}2016]\cite{bilgihan2016consumer}}
&\multicolumn{1}{p{2.2cm}}{\centering Structural equation modeling with a sample of travel-related OSN}
&\multicolumn{1}{p{1.5cm}}{\centering OSNs}
&\multicolumn{1}{p{2.2cm}}{\centering Confirmatory factor analysis}
&\multicolumn{1}{p{1.0cm}}{\centering Distributed}
&\multicolumn{1}{p{1.0cm}}{\centering \xmark}
&\multicolumn{1}{p{1.2cm}}{\centering \xmark}
&\multicolumn{1}{p{1cm}}{\centering \xmark}\\\\

\multicolumn{1}{p{2.2cm}}{\centering  Consumers relationship among elements of a brand community Model}
&\multicolumn{1}{p{2.2cm}}{\centering [Habibi \emph{et al.}2014]\cite{habibi2014roles}}
&\multicolumn{1}{p{2.2cm}}{\centering Based on brand, product, company, and other consumers}
&\multicolumn{1}{p{1.5cm}}{\centering OSNs}
&\multicolumn{1}{p{2.2cm}}{\centering Internal consistency, discriminant and convergent validity}
&\multicolumn{1}{p{1.0cm}}{\centering -}
&\multicolumn{1}{p{1.0cm}}{\centering \xmark}
&\multicolumn{1}{p{1.2cm}}{\centering \xmark}
&\multicolumn{1}{p{1cm}}{\centering \xmark}\\\\

\multicolumn{1}{p{2.2cm}}{\centering   Trust prediction strategy}
&\multicolumn{1}{p{2.2cm}}{\centering [Deni Raj and Babu 2017]\cite{raj2017enhanced}}
&\multicolumn{1}{p{2.2cm}}{\centering Probabilistic reputation features}
&\multicolumn{1}{p{1.5cm}}{\centering OSNs}
&\multicolumn{1}{p{2.2cm}}{\centering Accuracy comparison,F1 score}
&\multicolumn{1}{p{1.0cm}}{\centering -}
&\multicolumn{1}{p{1.0cm}}{\centering \xmark}
&\multicolumn{1}{p{1.2cm}}{\centering -}
&\multicolumn{1}{p{1cm}}{\centering \cmark}\\\\

\multicolumn{1}{p{2.2cm}}{\centering   Trustworthiness management}
&\multicolumn{1}{p{2.2cm}}{\centering [Nitti \emph{et al.}2014]\cite{nitti2014trustworthiness}}
&\multicolumn{1}{p{2.2cm}}{\centering Distributed hash table}
&\multicolumn{1}{p{1.5cm}}{\centering OSNs}
&\multicolumn{1}{p{2.2cm}}{\centering Transaction success rate, dynamic behavior}
&\multicolumn{1}{p{1.0cm}}{\centering  Distributed}
&\multicolumn{1}{p{1.0cm}}{\centering \xmark}
&\multicolumn{1}{p{1.2cm}}{\centering \xmark}
&\multicolumn{1}{p{1cm}}{\centering \cmark}\\\\

\multicolumn{1}{p{2.2cm}}{\centering   Secure PSN communications scheme}
&\multicolumn{1}{p{2.2cm}}{\centering [Yan \emph{et al.}2013]\cite{yan2013secure}}
&\multicolumn{1}{p{2.2cm}}{\centering Multi-dimensional trust levels}
&\multicolumn{1}{p{1.5cm}}{\centering OSNs}
&\multicolumn{1}{p{2.2cm}}{\centering Security Proofs, communication cost, computation complexity}
&\multicolumn{1}{p{1.0cm}}{\centering  Distributed}
&\multicolumn{1}{p{1.0cm}}{\centering \xmark}
&\multicolumn{1}{p{1.2cm}}{\centering \cmark}
&\multicolumn{1}{p{1cm}}{\centering \cmark}\\\\

\multicolumn{1}{p{2.2cm}}{\centering   Secure personal data access management scheme}
&\multicolumn{1}{p{2.2cm}}{\centering [Yan \emph{et al.} 2014]\cite{yan2014personal}}
&\multicolumn{1}{p{2.2cm}}{\centering Trust level with regard to a concrete context}
&\multicolumn{1}{p{1.5cm}}{\centering OSNs}
&\multicolumn{1}{p{2.2cm}}{\centering Data confidentiality, computation complexity}
&\multicolumn{1}{p{1.0cm}}{\centering  Distributed}
&\multicolumn{1}{p{1.0cm}}{\centering \xmark}
&\multicolumn{1}{p{1.2cm}}{\centering \xmark}
&\multicolumn{1}{p{1cm}}{\centering \cmark}\\\\

\multicolumn{1}{p{2.2cm}}{\centering  Anonymous authentication scheme}
&\multicolumn{1}{p{2.2cm}}{\centering [Yan \emph{et al.} 2015]\cite{yan2015anonymous}}
&\multicolumn{1}{p{2.2cm}}{\centering Batch-signature verification}
&\multicolumn{1}{p{1.5cm}}{\centering POSNs}
&\multicolumn{1}{p{2.2cm}}{\centering Operation time, communication cost}
&\multicolumn{1}{p{1.0cm}}{\centering  Centralized}
&\multicolumn{1}{p{1.0cm}}{\centering \xmark}
&\multicolumn{1}{p{1.2cm}}{\centering \xmark}
&\multicolumn{1}{p{1cm}}{\centering -}\\\\

\multicolumn{1}{p{2.2cm}}{\centering  Secure communication data}
&\multicolumn{1}{p{2.2cm}}{\centering [Huang \emph{et al.}2016]\cite{huang2016secure}}
&\multicolumn{1}{p{2.2cm}}{\centering Local trust evaluated by PSN nodes}
&\multicolumn{1}{p{1.5cm}}{\centering POSNs}
&\multicolumn{1}{p{2.2cm}}{\centering Generation time, operation time}
&\multicolumn{1}{p{1.0cm}}{\centering  Distributed}
&\multicolumn{1}{p{1.0cm}}{\centering \xmark}
&\multicolumn{1}{p{1.2cm}}{\centering \cmark}
&\multicolumn{1}{p{1cm}}{\centering \xmark}\\\\

\multicolumn{1}{p{2.2cm}}{\centering  Computational Offloading for Efficient Trust Management}
&\multicolumn{1}{p{2.2cm}}{\centering [Sharma \emph{et al.}2017]\cite{sharma2017computational}}
&\multicolumn{1}{p{2.2cm}}{\centering Osmotic Computing}
&\multicolumn{1}{p{1.5cm}}{\centering POSNs}
&\multicolumn{1}{p{2.2cm}}{\centering Osmosis time, computational overheads, relation cost}
&\multicolumn{1}{p{1.0cm}}{\centering  Distributed}
&\multicolumn{1}{p{1.0cm}}{\centering \xmark}
&\multicolumn{1}{p{1.2cm}}{\centering \cmark}
&\multicolumn{1}{p{1cm}}{\centering \cmark}\\\\

\multicolumn{1}{p{2.2cm}}{\centering  Trust evaluation scheme}
&\multicolumn{1}{p{2.2cm}}{\centering [Tajbakhsh \emph{et al.}2017]\cite{tajbakhsh2017computational}}
&\multicolumn{1}{p{2.2cm}}{\centering Three-valued subjective logic (3VSL)}
&\multicolumn{1}{p{1.5cm}}{\centering OSNs}
&\multicolumn{1}{p{2.2cm}}{\centering Probability of goodness, computational complexity}
&\multicolumn{1}{p{1.0cm}}{\centering  Distributed}
&\multicolumn{1}{p{1.0cm}}{\centering \xmark}
&\multicolumn{1}{p{1.2cm}}{\centering \xmark}
&\multicolumn{1}{p{1cm}}{\centering \cmark}\\\\
\hline
\end{longtable}
\end{center}
\end{landscape}

\begin{figure}[ht!]
\centering
\includegraphics[width=300px]{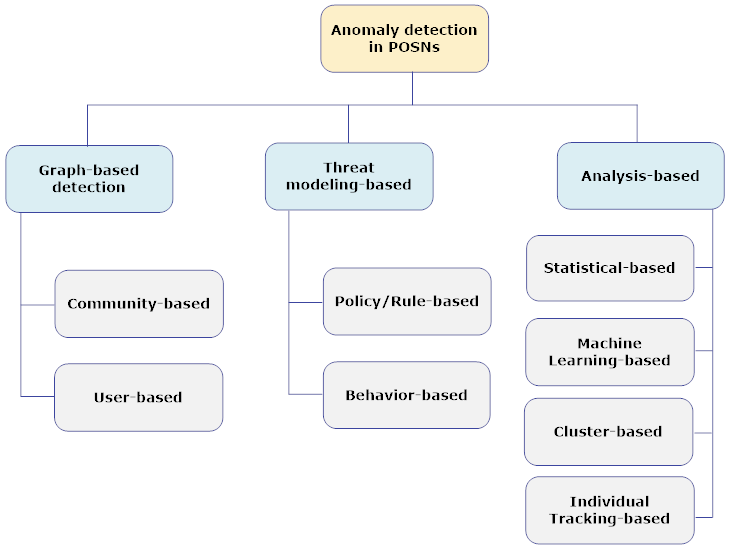}
\caption{A taxonomy of anomaly detection approaches in POSNs.}
\label{tax_ano}
\end{figure}


\section{Anomaly detection in POSNs}
\label{sect:Taxonomy_Anomaly}
The anomalies are the illegal or unwanted behavior in OSNs that show harmful effects on the cross-platform POSNs~\cite{viswanath2014towards}~\cite{sharma2017computational}. Therefore, the solutions against anomaly detection are required in POSNs. The anomalies can be categorized into following parts, as shown in Figure~\ref{tax_ano}:
\begin{itemize}
  \item \textbf{Point Anomalies :} The anomalies over independent data instance is known as point anomalies. For example, in POSNs, the user tries to access a restricted service and leads to different kind of attacks. Such an anomaly is the resultant of an individual's activity~\cite{kaur2016survey}~\cite{savage2014anomaly}~\cite{anand2017anomaly}.
  \item \textbf{Contextual Anomalies :} The anomalies over a specific dataset with a specific context like time, region, etc., are known as contextual anomalies~\cite{kaur2016survey}~\cite{savage2014anomaly}~\cite{anand2017anomaly}.
  \item \textbf{Collective Anomalies :} The anomalies of complete data set is considered as collective anomalies. These anomalies can be detected only when an event occurs in an unexpected order or with an unexpected combination of values or users~\cite{kaur2016survey}~\cite{savage2014anomaly}~\cite{anand2017anomaly}.
\item \textbf{Horizontal Anomalies :} The anomalies due to the difference in the activity and interaction policies of a user with different sources are categorized as horizontal anomalies. Such anomalies are most difficult and tedious to detect as an error in detection may lead to high values for false positives as well as false negatives~\cite{8322278}~\cite{kaur2016survey}~\cite{savage2014anomaly}~\cite{anand2017anomaly}.
\end{itemize}

On the basis of nature of data and behavior of users or communities, various anomaly detection techniques can be used to detect an anomaly in POSNs as explained below:

\subsection{Graph-based anomaly detection}
Some users would have different types of interactions with different communities. Such social relationship among friends can be used to create dependency graph among participants and these graphs can be used to detect anomalies~\cite{noble2003graph}~\cite{akoglu2015graph}. These graphs help in finding unexpected behavior or any user or community that shows an abnormal activity. Graph-based anomaly detection can be further categorized into two parts,

\subsubsection{Community-based}
In such type, the graphs are used to detect abnormal behavior of any community. The dependency graphs are created among communities; therefore, the collected data is large and include many communities~\cite{8322278}~\cite{sharma2017computational}. The contextual and collective anomalies in POSNs can be detected through these graphs.

\subsubsection{User-based}
In this approach, an individual is targeted to detect anomalies. The dependency graph of individual users is used as a data set to find a suspected node or faulty user in the relationships~\cite{sharma2017isma}.

\subsection{Threat-modeling based anomaly detection}
The threat is any suspicious or abnormal activity which causes harmful effects or exploits on POSNs. The threat modeling helps to find any type of anomalies in POSNs. The threat modeling is performed with the help of various parameters like behaviors, the flow of information among users, the characteristics of users and communities~\cite{myagmar2005threat}~\cite{grossman2018simulation}. Threat-modeling based anomaly detection can be further classified into two types:

\subsubsection{Policy/Rule-based}
In the policies, the threat modeling is done with the help of characteristics and flow of information over POSNs~\cite{wang2017spoke}. The policies or rules decide conditions to find anomalies which show any abnormality over these rules.
\subsubsection{Behavior-based}
The behavior of user or communities is used as a model for threat detection in POSNs. The threat is considered as bad behavior by users and listed as a bad indicator in the model~\cite{sharma2017driver}. When any of the bad behaviors are reflected in the data set, anomalies are detected instantly.
\subsection{Analysis based anomaly detection}
The information in POSNs is stored in the form of datasets. The anomalies are detected on these datasets with the help of analysis and data mining either by individual or third party operators~\cite{aljawarneh2017anomaly}~\cite{zhang2018convergence}. Moreover, the analysis can be performed dynamically or statically over the POSN datasets. Analysis based anomaly detection can be broadly classified into following types:

\subsubsection{Statistical-based}
The simplest approach of data analysis to find an anomaly is statistical analysis. In this type, various common statistical properties of distributions, including mean, median, and mode are used to identify irregularities in data. The statistical analysis based filters are good at anomaly detections but for highly uncertain data sets, these may generate false information~\cite{bigdeli2017fast}. The noise among users is also considered as an abnormal behavior and generated as false anomalies.
\begin{landscape}
\begin{center}
\fontsize{7}{10}\selectfont
\setlength\LTleft{30pt}            
\setlength\LTright{0pt}
\begin{longtable}{@{\extracolsep{\fill}}*{8}{c}}
\caption{State-of-the-art anomaly detection solutions for POSNs.}\label{Table1} \\
\hline
\multicolumn{1}{p{2.2cm}}{\centering \textbf{Approach}}
&\multicolumn{1}{p{2.2cm}}{\centering \textbf{Author} }
&\multicolumn{1}{p{2.2cm}}{\centering \textbf{Ideology}}
&\multicolumn{1}{p{2cm}}{\centering \textbf{Application} }
&\multicolumn{1}{p{2.2cm}}{\centering \textbf{Parameter}}
&\multicolumn{1}{p{1.2cm}}{\centering \textbf{Mechanism}}
&\multicolumn{1}{p{1.0cm}}{\centering \textbf{Threat}\\ \textbf{Modeling}}
&\multicolumn{1}{p{1.0cm}}{\centering \textbf{Recovery}}\\[6pt]\\
\hline \\

\endfirsthead

\multicolumn{8}{c}%
{{\bfseries \tablename\ \thetable{} -- continued from previous page}} \\
\hline

\multicolumn{1}{p{2.2cm}}{\centering \textbf{Approach}}
&\multicolumn{1}{p{2.2cm}}{\centering \textbf{Author} }
&\multicolumn{1}{p{2.2cm}}{\centering \textbf{Ideology}}
&\multicolumn{1}{p{2cm}}{\centering \textbf{Application} }
&\multicolumn{1}{p{2.2cm}}{\centering \textbf{Parameter}}
&\multicolumn{1}{p{1.2cm}}{\centering \textbf{Mechanism}}
&\multicolumn{1}{p{1.0cm}}{\centering \textbf{Threat}\\ \textbf{Modeling}}
&\multicolumn{1}{p{1.0cm}}{\centering \textbf{Recovery}}\\[6pt]\\
\hline\\\endhead

\hline \multicolumn{8}{l}{{Continued on next page}} \\

\endfoot

\endlastfoot

\multicolumn{1}{p{2.2cm}}{\centering  Tension detection approach in online communities}
&\multicolumn{1}{p{2.2cm}}{\centering [Burnap \emph{et al.} 2015]\cite{burnap2015detecting}}
&\multicolumn{1}{p{2.2cm}}{\centering Conversation analysis combined with syntactic and lexicon-based text mining rules}
&\multicolumn{1}{p{2cm}}{\centering OSNs}
&\multicolumn{1}{p{2.2cm}}{\centering Sentiment analysis, machine learning}
&\multicolumn{1}{p{1.2cm}}{\centering On-site}
&\multicolumn{1}{p{1.0cm}}{\centering \xmark}
&\multicolumn{1}{p{1.0cm}}{\centering \xmark} \\[6pt]\\

\multicolumn{1}{p{2.2cm}}{\centering  Anomaly detection mechanisms}
&\multicolumn{1}{p{2.2cm}}{\centering [Garroppo and Niccolini 2018]\cite{garroppo2018anomaly}}
&\multicolumn{1}{p{2.2cm}}{\centering Spatial analysis (SA) and time analysis (TA)}
&\multicolumn{1}{p{2cm}}{\centering Cellular systems}
&\multicolumn{1}{p{2.2cm}}{\centering Traffic data and alarms, behaviour, artificial anomalies}
&\multicolumn{1}{p{1.2cm}}{\centering On-site}
&\multicolumn{1}{p{1.0cm}}{\centering \xmark}
&\multicolumn{1}{p{1cm}}{\centering \xmark} \\[6pt]\\

\multicolumn{1}{p{2.2cm}}{\centering  Anomaly detection}
&\multicolumn{1}{p{2.2cm}}{\centering [Savage \emph{et al.} 2014]\cite{savage2014anomaly}}
&\multicolumn{1}{p{2.2cm}}{\centering The selection and calculation of network features, and the classification of observations from this feature space}
&\multicolumn{1}{p{2cm}}{\centering OSNs}
&\multicolumn{1}{p{2.2cm}}{\centering Labelled and unlabelled anomalies}
&\multicolumn{1}{p{1.2cm}}{\centering -}
&\multicolumn{1}{p{1cm}}{\centering \xmark}
&\multicolumn{1}{p{1cm}}{\centering \xmark} \\[6pt]\\

\multicolumn{1}{p{2.2cm}}{\centering OLAIR- behaviors detection}
&\multicolumn{1}{p{2.2cm}}{\centering [Wang and Cheng 2005]\cite{wang2005detecting}}
&\multicolumn{1}{p{2.2cm}}{\centering Social network analysis}
&\multicolumn{1}{p{2cm}}{\centering OSNs}
&\multicolumn{1}{p{2.2cm}}{\centering SNA indicator, k-core value}
&\multicolumn{1}{p{1.2cm}}{\centering Off-site}
&\multicolumn{1}{p{1cm}}{\centering \xmark}
&\multicolumn{1}{p{1cm}}{\centering \xmark} \\[6pt]\\

\multicolumn{1}{p{2.2cm}}{\centering Utility cloud anomalies detection}
&\multicolumn{1}{p{2.2cm}}{\centering [Wang \emph{et al.} 2010]\cite{wang2010online}}
&\multicolumn{1}{p{2.2cm}}{\centering Metric distributions}
&\multicolumn{1}{p{2cm}}{\centering OSNs}
&\multicolumn{1}{p{2.2cm}}{\centering Baseline methods, entropy time}
&\multicolumn{1}{p{1.2cm}}{\centering On-site}
&\multicolumn{1}{p{1cm}}{\centering \xmark}
&\multicolumn{1}{p{1cm}}{\centering \xmark} \\[6pt]\\

\multicolumn{1}{p{2.2cm}}{\centering Anomaly detection with system feedback}
&\multicolumn{1}{p{2.2cm}}{\centering [Horn and Willett 2011]\cite{horn2011online}}
&\multicolumn{1}{p{2.2cm}}{\centering Filtering and hedging}
&\multicolumn{1}{p{2cm}}{\centering OSNs}
&\multicolumn{1}{p{2.2cm}}{\centering Detection misses, false alarms correct anomalies}
&\multicolumn{1}{p{1.2cm}}{\centering On-site}
&\multicolumn{1}{p{1cm}}{\centering \xmark}
&\multicolumn{1}{p{1cm}}{\centering \xmark} \\[6pt]\\

\multicolumn{1}{p{2.2cm}}{\centering Eigenspace Analysis for threat detection}
&\multicolumn{1}{p{2.2cm}}{\centering [Miller \emph{et al.} 2011]\cite{miller2011eigenspace}}
&\multicolumn{1}{p{2.2cm}}{\centering Detection of a threat subgraph}
&\multicolumn{1}{p{2cm}}{\centering OSNs}
&\multicolumn{1}{p{2.2cm}}{\centering Threat detection performance}
&\multicolumn{1}{p{1.2cm}}{\centering Off-site}
&\multicolumn{1}{p{1cm}}{\centering \cmark}
&\multicolumn{1}{p{1cm}}{\centering \xmark} \\[6pt]\\

\multicolumn{1}{p{2.2cm}}{\centering SNAD}
&\multicolumn{1}{p{2.2cm}}{\centering [Chen \emph{et al.} 2011]\cite{chen2011leveraging}}
&\multicolumn{1}{p{2.2cm}}{\centering Assembles the community of users that access a particular subject and assesses if similarities of the community}
&\multicolumn{1}{p{2cm}}{\centering OSNs}
&\multicolumn{1}{p{2.2cm}}{\centering Detection performance}
&\multicolumn{1}{p{1.2cm}}{\centering Off-site}
&\multicolumn{1}{p{1cm}}{\centering \cmark}
&\multicolumn{1}{p{1cm}}{\centering \xmark} \\[6pt]\\

\multicolumn{1}{p{2.2cm}}{\centering Link anomaly detection}
&\multicolumn{1}{p{2.2cm}}{\centering [Takahashi \emph{et al.}2014]\cite{takahashi2014discovering}}
&\multicolumn{1}{p{2.2cm}}{\centering Sequentially discounting normalized maximum likelihood (SDNML) and with Kleinberg’s burst model}
&\multicolumn{1}{p{2cm}}{\centering OSNs}
&\multicolumn{1}{p{2.2cm}}{\centering Dynamic threshold optimization, probability modeling}
&\multicolumn{1}{p{1.2cm}}{\centering On-site}
&\multicolumn{1}{p{1cm}}{\centering \xmark}
&\multicolumn{1}{p{1cm}}{\centering \xmark} \\[6pt]\\

\multicolumn{1}{p{2.2cm}}{\centering Proactive insider threat detection}
&\multicolumn{1}{p{2.2cm}}{\centering [Brdiczka \emph{et al.} 2012]\cite{brdiczka2012proactive}}
&\multicolumn{1}{p{2.2cm}}{\centering Graph learning and psychological context}
&\multicolumn{1}{p{2cm}}{\centering Online Games}
&\multicolumn{1}{p{2.2cm}}{\centering Personality predictions, network statistics}
&\multicolumn{1}{p{1.2cm}}{\centering On-site}
&\multicolumn{1}{p{1cm}}{\centering \cmark}
&\multicolumn{1}{p{1cm}}{\centering \xmark} \\[6pt]\\

\multicolumn{1}{p{2.2cm}}{\centering iBOAT}
&\multicolumn{1}{p{2.2cm}}{\centering [Chen \emph{et al.}2013]\cite{chen2013iboat}}
&\multicolumn{1}{p{2.2cm}}{\centering Isolation-based}
&\multicolumn{1}{p{2cm}}{\centering GPS}
&\multicolumn{1}{p{2.2cm}}{\centering Weighting function, anomaly score}
&\multicolumn{1}{p{1.2cm}}{\centering Off-site}
&\multicolumn{1}{p{1cm}}{\centering \xmark}
&\multicolumn{1}{p{1cm}}{\centering \xmark} \\[6pt]\\

\multicolumn{1}{p{2.2cm}}{\centering Rule-based hybrid anomaly detection method}
&\multicolumn{1}{p{2.2cm}}{\centering [Hassanzadeh and Nayak 2013]\cite{hassanzadeh2013rule}}
&\multicolumn{1}{p{2.2cm}}{\centering Graph theory, Fuzzy clustering and Fuzzy rules for modeling user relationships}
&\multicolumn{1}{p{2cm}}{\centering OSNs}
&\multicolumn{1}{p{2.2cm}}{\centering Fuzzy behaviors, graph metrics}
&\multicolumn{1}{p{1.2cm}}{\centering Off-site}
&\multicolumn{1}{p{1cm}}{\centering \xmark}
&\multicolumn{1}{p{1cm}}{\centering \xmark} \\[6pt]\\

\multicolumn{1}{p{2.2cm}}{\centering Anomaly detection method}
&\multicolumn{1}{p{2.2cm}}{\centering  [Rezaei \emph{et al.}2013]\cite{rezaei2013anomaly}}
&\multicolumn{1}{p{2.2cm}}{\centering Structure-based technique}
&\multicolumn{1}{p{2cm}}{\centering OSNs}
&\multicolumn{1}{p{2.2cm}}{\centering Fitting Curve, graph metrics, anomaly score}
&\multicolumn{1}{p{1.2cm}}{\centering Off-site}
&\multicolumn{1}{p{1cm}}{\centering \xmark}
&\multicolumn{1}{p{1cm}}{\centering \xmark} \\[6pt]\\

\multicolumn{1}{p{2.2cm}}{\centering TargetVue: visual analysis system}
&\multicolumn{1}{p{2.2cm}}{\centering  [Cao \emph{et al.}2016]\cite{cao2016targetvue}}
&\multicolumn{1}{p{2.2cm}}{\centering Unsupervised learning model-visualizes the behaviors of suspicious users}
&\multicolumn{1}{p{2cm}}{\centering Online Communication Systems}
&\multicolumn{1}{p{2.2cm}}{\centering Relation glyph, Z-glyph, behavior glyph}
&\multicolumn{1}{p{1.2cm}}{\centering Off-site}
&\multicolumn{1}{p{1cm}}{\centering \cmark}
&\multicolumn{1}{p{1cm}}{\centering \xmark} \\[6pt]\\

\multicolumn{1}{p{2.2cm}}{\centering COMPA: Compromised accounts detection method}
&\multicolumn{1}{p{2.2cm}}{\centering  [Egele \emph{et al.}2017]\cite{egele2017towards}}
&\multicolumn{1}{p{2.2cm}}{\centering Based on a simple observation: social network users develop habits over time, and these habits are fairly stable}
&\multicolumn{1}{p{2cm}}{\centering OSNs}
&\multicolumn{1}{p{2.2cm}}{\centering Behavioral profile stability, False Positives and negative }
&\multicolumn{1}{p{1.2cm}}{\centering On-site}
&\multicolumn{1}{p{1cm}}{\centering \cmark}
&\multicolumn{1}{p{1cm}}{\centering \xmark} \\[6pt]\\

\multicolumn{1}{p{2.2cm}}{\centering Voila }
&\multicolumn{1}{p{2.2cm}}{\centering  [Cao \emph{et al.}2018]\cite{cao2018voila}}
&\multicolumn{1}{p{2.2cm}}{\centering Interactively detecting anomalies in spatiotemporal data collected from a streaming data source}
&\multicolumn{1}{p{2cm}}{\centering Spatiotemporal data}
&\multicolumn{1}{p{2.2cm}}{\centering Ground-truth labeling, baseline methods and evaluation metrics }
&\multicolumn{1}{p{1.2cm}}{\centering On-site}
&\multicolumn{1}{p{1cm}}{\centering \xmark}
&\multicolumn{1}{p{1cm}}{\centering \xmark} \\[6pt]\\

\multicolumn{1}{p{2.2cm}}{\centering COSMOS }
&\multicolumn{1}{p{2.2cm}}{\centering  [Burnap \emph{et al.}2015]\cite{burnap2015cosmos}}
&\multicolumn{1}{p{2.2cm}}{\centering Hadoop infrastructure}
&\multicolumn{1}{p{2cm}}{\centering OSNs}
&\multicolumn{1}{p{2.2cm}}{\centering Sentiment analysis, social network analysis, hadoop scalability }
&\multicolumn{1}{p{1.2cm}}{\centering Off-site}
&\multicolumn{1}{p{1cm}}{\centering \xmark}
&\multicolumn{1}{p{1cm}}{\centering \xmark} \\[6pt]\\

\multicolumn{1}{p{2.2cm}}{\centering Bayesian anomaly detection methods }
&\multicolumn{1}{p{2.2cm}}{\centering  [Heard \emph{et al.}2010]\cite{heard2010bayesian}}
&\multicolumn{1}{p{2.2cm}}{\centering Track the pair wise links of all nodes in the graph to assess normality of behavior}
&\multicolumn{1}{p{2cm}}{\centering OSNs}
&\multicolumn{1}{p{2.2cm}}{\centering Sequential and retrospective analyses }
&\multicolumn{1}{p{1.2cm}}{\centering On-site}
&\multicolumn{1}{p{1cm}}{\centering \xmark}
&\multicolumn{1}{p{1cm}}{\centering \xmark} \\[6pt]\\

\multicolumn{1}{p{2.2cm}}{\centering Bursty keyword detection model }
&\multicolumn{1}{p{2.2cm}}{\centering  [Guzman and Poblete 2013]\cite{guzman2013line}}
&\multicolumn{1}{p{2.2cm}}{\centering Normalized individual frequency signals per term and a windowing variation technique}
&\multicolumn{1}{p{2cm}}{\centering OSNs}
&\multicolumn{1}{p{2.2cm}}{\centering Stopword analysis, Scalability }
&\multicolumn{1}{p{1.2cm}}{\centering Off-site}
&\multicolumn{1}{p{1cm}}{\centering \xmark}
&\multicolumn{1}{p{1cm}}{\centering \xmark} \\[6pt]\\

\multicolumn{1}{p{2.2cm}}{\centering CatchSync }
&\multicolumn{1}{p{2.2cm}}{\centering  [Jiang  \emph{et al.}2014]\cite{jiang2014detecting}}
&\multicolumn{1}{p{2.2cm}}{\centering Synchronized behavior and abnormal behavior}
&\multicolumn{1}{p{2cm}}{\centering OSNs}
&\multicolumn{1}{p{2.2cm}}{\centering Detection effectiveness }
&\multicolumn{1}{p{1.2cm}}{\centering Off-site}
&\multicolumn{1}{p{1cm}}{\centering \xmark}
&\multicolumn{1}{p{1cm}}{\centering \xmark} \\[6pt]\\

\hline
\end{longtable}
\end{center}
\end{landscape}

\subsubsection{Machine learning-based}
The instant anomalies like zero-day attacks or unexploited vulnerabilities are not detected through statistical analysis approaches. Machine learning-based analysis is more effective in such cases. The machine learning approach includes k-nearest neighbors’ algorithm, local outlier factor and supervised learning through vector machine approaches. However, the machine learning solutions are quite expensive to maintain because of additional overheads~\cite{faigon2017machine}~\cite{gao2018cloud}.

\subsubsection{Cluster-based}
The clusters are composed of users and communities with similar properties~\cite{brandsaeter2017cluster}. On the basis of these common properties, the clusters are created and analyzed for anomaly detections.

\subsubsection{Individual tracking-based}
The individual group or any user is traced for anomalies detection through this approach~\cite{glas2017personal}. In the individual tracking, every user characteristics are analyzed to obtain any misbehavior in POSNs.

A detailed comparison of existing solutions on the anomaly detection in POSNs is presented in Table~\ref{Table1}.

\section{Research Challenges and Future Directions}
\label{sect:roadmap}
POSNs are in the growing stage as the amalgamation of different OSNs has already started. The rapid growth of such networks causes a considerable impact on the research organizations and demand solutions for different types of issues discussed throughout this article. In order to facilitate the future research, different open issues research challenges are discussed below:

\begin{itemize}
\item \textbf{Anonymous Authentication :} The possibility of user communication in POSNs needs to judge whether the communication parties are trustworthy or not. The concept of anonymous authenticating supports trust of the involved parties while preserving their privacy in POSNs~\cite{farash2017lightweight}~\cite{gao2017anonymous}. It further involves extra computational overheads in terms of operational time and communication costs. The future work should be directed towards the security paradigms of POSNs and should incorporate reliable privacy preservation via trust, anonymity, and unlinkability in the nodes, protect traceability and link predictions through low computational anonymous authentication mechanisms.

\item \textbf{Inter-vehicular POSNs:} With the advent of vehicular social networks, this field becomes important to cover from POSNs perspective. At the moment, POSNs through vehicular technology has not been studied much and requires some considerable evaluations on demonstrating how V2X can be introduced into the existing setup while showcasing its utility as one of the platforms~\cite{sharma2016energy}. Moreover, security aspects are entirely open in this direction of research and it can serve as an important and highly motivated topic for upcoming research. Integration of different technologies like crowdsourcing, the blockchain, osmotic computing, catalytic computing can be performed to use vehicles as one of the nodes in POSNs~\cite{sharma2018securea,rahim2017vehicular,fantacci2018bio,sharma2017socializing,sharma2017resource}. In addition, mobility-aware data sharing is also an open issue for inter-vehicular POSNs.

\item \textbf{Mutual Trust Management :} Trust is considered as an important asset or a property of relationships for networks that help to shape interaction patterns within POSNs. In general, the trust is established among two participants who have direct interactions in a large-scale social network. Mutual trust is required between the communicating parties to ensure each other and verify the source of information~\cite{de2017combining,de2017forming,jowua17-8-3-01,jowua17-8-4-04}. Different structural and relational properties of users in POSNs need mutual trust amongst its users. The further considerable issues in POSNs include level-wise mutual trust and identification of factors which influence trust.

\item \textbf{Light-Weight Security :} POSNs contain a wealth of information about its users embedded in the social graph and links. It is important to develop new lightweight cryptographic algorithms which protect the privacy and security of communication between the communities as well as users. The lightweight security requires developing cryptographic algorithms and standards that can work within the confines of a simple POSN network and users~\cite{ma2018armor,sharma2018secure,zhang2017population,jisis17-7-4-02}. In POSNs, lightweight security algorithm, protocol construction, and implementation are still open issues to resolve.

\item \textbf{Cross-platform POSNs Management :} POSNs require combined efforts from different platforms to manage its users especially focusing on their security considerations. It is required that low-complex and low-cost solutions should be developed which can facilitate the easier management of multiple platforms of POSNs. Such requirements become severe when the platforms are independent of each other and have the least number of metrics in common~\cite{sharma2017computational,li2017privacy,li2016secure}. This condition makes it difficult to manage and control the activities of POSNs. Research is required in this direction and it is expected that the developed solutions can be implemented without getting affecting of the cross-platform data exchanges.

\item \textbf{Secure Recommender Systems:} The recommender systems provide personalized information to users according to their preferences like traveling data, shopping lists, tourism, etc.  It is desired that recommender systems which target the audience of POSNs must be secure and should not allow security breaches to prevent manipulation of its users. Recommender systems can be strengthened for data privacy and can use various authentication mechanisms for the end to end security~\cite{shmueli2017secure}~\cite{zhang2018privacy}.

\item \textbf{Access Control and Authorization :} The content over POSNs needs to access policies and authorization through permissions. To prevent unwanted access to the content, access control and user-specified policies are required that helps to regulate how a user accesses and controls the information. Some of the future challenges involve multiparty access control in POSNs, relationship-based access controls, and secondary authorizations~\cite{sharma2017conf1,levergood2018internet,park2017role,sharma2017conf2}.

\item \textbf{Resilience to Failures :} Nodes failures are difficult to manage for any kind of networks. However, such an issue becomes critical if the primary node which is an interface between two or more platforms fails. For cross-platform data exchanges in POSNs, it is required that either the system should be made fail-safe, or the network must be checked for a single point of failures~\cite{woods2017prologue,7268858,zhao2017analyzing,sharma2018lorawan}. Thus, it is desirable to develop platforms that are resilient to periodic failures.

\item \textbf{Secure Resource Allocation :} The resources should be efficiently divided amongst POSN users to fulfill the need of an individual without leading to resource-starvations. Owners of the resources must share their computing resources for their friend's circle in a secure manner~\cite{sharma2018self}. At a glance, the distributed resources and infrastructures are shared with the communities in POSNs. Therefore, the secure resource allocation is a considerable challenge to resolve while leveraging the facilities of present solutions. From the research perspective, distributed resource allocations, preference-based resource sharing, and Infrastructure sharing in POSNs need a considerable attention~\cite{su2017security,shila2017amcloud,7933191}.

\item \textbf{POSNs Survivability :} The survivability of POSNs can be related to its resource allocation and resilience to failures. Accountability of both these factors helps to make POSNs survivable as well as sustainable irrespective of the functioning conditions and environment~\cite{li2017improving,8000322,armitage2017communities}. POSNs are focused on supporting a large set of daily activities, thus, it is required that frameworks and middleware should be developed that can ensure the high survivability of POSNs.
\end{itemize}
\begin{figure}
  \centering
  \includegraphics[width=370px]{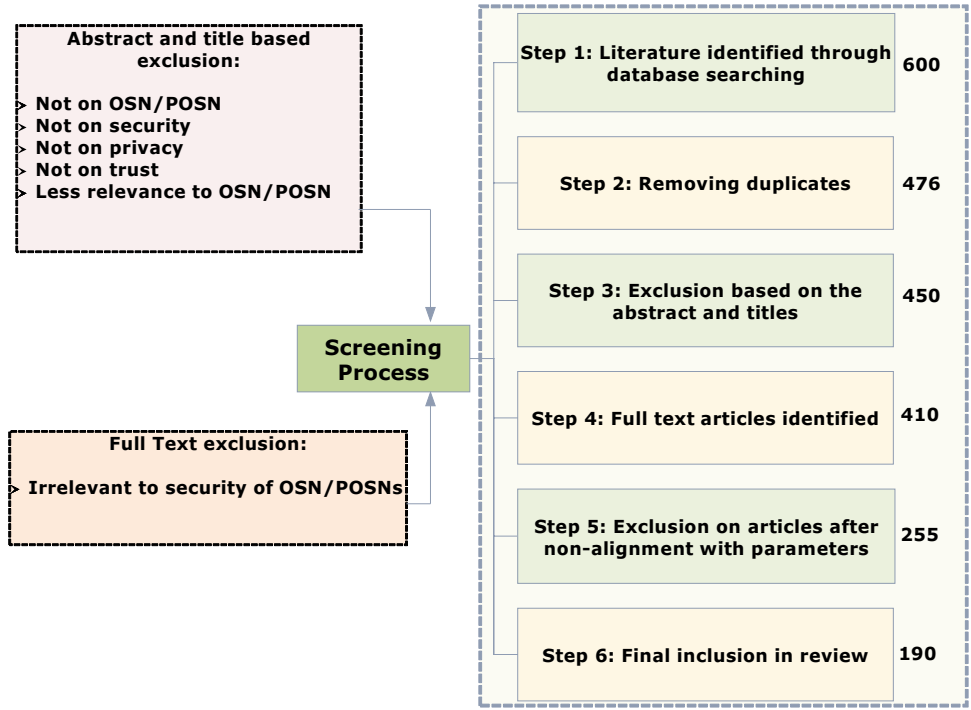}
  \caption{An illustration of the screening strategy opted for this survey.}\label{fig9a}
\end{figure}

\begin{figure}
  \centering
  \includegraphics[width=330px]{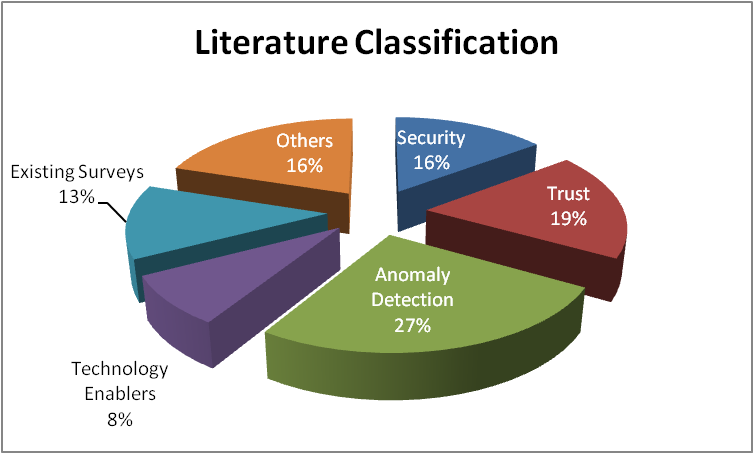}
  \caption{Literature classification for comparative evaluations of 190 articles.}\label{fig7a}
\end{figure}
\begin{figure}
  \centering
  \includegraphics[width=340px]{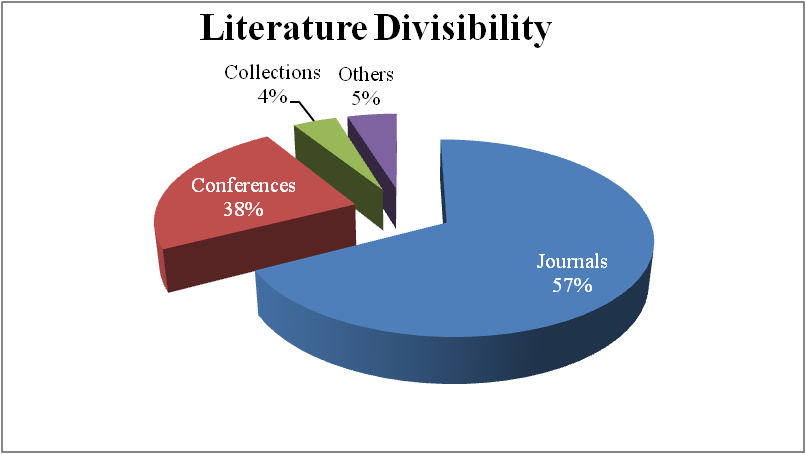}
  \caption{Literature divisibility of 190 articles.}\label{fig8a}
\end{figure}

\section{Literature Classification}
\label{sect:Literature_Classification}
The literature was extracted from the online database searches using Google Scholar, IEEE Xplore, ACM Digital libraries, Web of Science directory, Science Direct, university library search, and backtracking the references of existing articles. The procedure followed for identification of relevant material is illustrated in Figure~\ref{fig9a}. Figures~\ref{fig7a} and~\ref{fig8a} illustrate the literature classification and divisibility, respectively, for different metrics for a total of 190 articles, which are referred to in this survey.
\section{Conclusions}
\label{sect:Conclusions}
Pervasive Online Social Networks (POSNs) aim at connecting users from different locations, domains, platforms and with different properties of relationships. POSNs focus on cross-platform communications amongst large numbers of users in the form of communities belonging to two or more OSN platforms. Such formations are difficult to handle and security is one of the crucial challenges to tackle for attaining a fully-flexible workflow model for POSNs. At present, there are limited information and contents on the security of POSNs. Irrespective of this, different sets of solutions which emphasize on social networking are studied and a detailed classification is presented for enhancing the knowledge about such solutions. In addition, this survey included content related to trust management and anomaly detection in POSNs. A broad classification is also presented for each category with a tabular comparison. Future challenges, open issues, and research goals are presented to provide a direction of research to the upcoming researchers. A total of 190 articles from different journals, conference-proceedings, and collections are referred with the majority of them focusing on closely related aspects of OSNs and POSNs. The understanding of this literature suggests that although there has been a tremendous amount of works on the security of OSNs, the challenge becomes crucial when cross-platform security is involved which is the dominant aspect of POSNs. Considering this, there is a huge amount of scope and gap that has to be filled with efficient and effective strategies.



%
\label{sect:bib}
\bibliographystyle{abbrv}
\bibliography{related_works}
\section*{Author Biography}
\vspace*{4em}
\begin{biography}{Takshi Gupta}{a1} received the B.Tech. degree in Computer Science and Engineering from Punjab Technical University in 2012 and the PG Diploma in Business Administration from SCDL, Symbiosis University, India in 2014 both with distinction. She is currently associated with the MobiSec Lab, Department of Information Security Engineering, Soonchunhyang University, Asan, South Korea. Prior to this, she worked at Kochar InfoTech followed by Gen-XT Infosystems and was a co-founder at Future Info-Systems. Her areas of research and interests are database management and security, artificial intelligence, semantic webs, and human resource management.
\end{biography}
\vspace*{2em}
\begin{biography}{Gaurav Choudhary}{a2} received the B.Tech. degree in Computer Science and Engineering from Rajasthan Technical University in 2014 and the Master Degree in Cyber Security from Sardar Patel University of Police in 2017. He is currently pursuing Ph.D. degree in the Department of Information Security Engineering, Soonchunhyang University, Asan, South Korea. His areas of research and interests are UAVs, Mobile and Internet security, IoT security, Network security, and Cryptography.
\\
\\
\end{biography}
\vspace*{2em}
\begin{biography}{Vishal Sharma}{a3} received the Ph.D. and B.Tech. degrees in computer science and engineering from Thapar University (2016) and Punjab Technical University (2012), respectively. He worked at Thapar University as a Lecturer from Apr'16-Oct'16. From Nov. 2016 to Sept. 2017, he was a joint post-doctoral researcher in MobiSec Lab. at Department of Information Security Engineering, Soonchunhyang University, and Soongsil University, Republic of Korea. Dr. Sharma is now a Research Assistant Professor in the Department of Information Security Engineering, Soonchunhyang University, The Republic of Korea. Dr. Sharma received three best paper awards from the IEEE International Conference on Communication, Management and Information Technology (ICCMIT), Warsaw, Poland in April 2017; from CISC-S'17 South Korea in June 2017; and from IoTaas Taiwan in September 2017. He is the member of IEEE, a professional member of ACM and past Chair for ACM Student Chapter-TU Patiala. He has authored/coauthored more than 60 journal/conference articles and bookchapters. He serves as the program committee member for the Journal of Wireless Mobile Networks, Ubiquitous Computing, and Dependable Applications (JoWUA). He was the track chair of MobiSec'16 and AIMS-FSS'16, and PC member and reviewer of MIST'16 and MIST'17, respectively. He was the TPC member of ITNAC-IEEE TCBD'17 and serving as TPC member of ICCMIT'18, CoCoNet'18 and ITNAC-IEEE TCBD'18. Also, he serves as a reviewer for various IEEE Transactions and other journals. His areas of research and interests are 5G networks, UAVs, estimation theory, and artificial intelligence.
\end{biography}
\end{document}